\documentclass[%
 aip,
 amsmath,amssymb,
 reprint,%
]{revtex4-1}

\usepackage{graphicx}
\usepackage{dcolumn}
\usepackage{bm}

\usepackage[utf8]{inputenc}
\usepackage[T1]{fontenc}
\usepackage{mathptmx}
\usepackage{etoolbox}
\usepackage[svgnames]{xcolor}
\usepackage{placeins}
\usepackage{enumerate}
\usepackage{hyperref}
\hypersetup{
    colorlinks,
    citecolor=blue,
    linkcolor=blue,
    filecolor=blue,      
    urlcolor=blue,
    pdftitle=blue,
    pdfpagemode=FullScreen,
    }
\usepackage[nameinlink]{cleveref}
\usepackage{todonotes}

\crefname{Appendix}{Appendix}{Appendices}

\makeatletter
\def\@email#1#2{%
 \endgroup
 \patchcmd{\titleblock@produce}
  {\frontmatter@RRAPformat}
  {\frontmatter@RRAPformat{\produce@RRAP{*#1\href{mailto:#2}{#2}}}\frontmatter@RRAPformat}
  {}{}
}%
\makeatother
\begin{document}

\preprint{AIP/123-QED}

\title{The climatic interdependence of extreme-rainfall events around the globe}
\author{Zhen Su}
\email{zhen.su@pik-potsdam.de}
\affiliation{Potsdam Institute for Climate Impact Research, 14473 Potsdam, Germany}
\affiliation{Department of Computer Science, Humboldt-Universität zu Berlin, 10099 Berlin, Germany}

\author{Henning Meyerhenke}
\affiliation{Department of Computer Science, Humboldt-Universität zu Berlin, 10099 Berlin, Germany}

\author{Jürgen Kurths}
\affiliation{Potsdam Institute for Climate Impact Research, 14473 Potsdam, Germany}
\affiliation{Department of Physics, Humboldt-Universität zu Berlin, 10099 Berlin, Germany}

\date{\today}

\begin{abstract}
    The identification of regions of similar climatological behavior can be utilized for the discovery of spatial relationships over long-range scales, including teleconnections.
    Additionally, it provides insights for the improvement of corresponding interaction processes in {general circulation models (GCMs).}
    In this regard, the global picture of the interdependence patterns of extreme rainfall events (EREs) still needs to be further explored.
    To this end, we propose a top-down complex-network-based clustering workflow, {with the combination of consensus clustering and mutual correspondences.
            Consensus clustering provides a reliable community structure under each dataset, while mutual correspondences build a matching relationship between different community structures obtained from different datasets.
            This approach ensures the robustness of the identified structures when multiple datasets are available.}
    By applying it simultaneously to two satellite-derived precipitation datasets, we identify consistent synchronized structures of EREs around the globe, during boreal summer.
    Two of them show independent spatiotemporal characteristics, uncovering the primary compositions of different monsoon systems.
    They explicitly manifest the primary intraseasonal variability in the context of the global monsoon, in particular the `monsoon jump' over both East Asia and West Africa and the mid-summer drought over Central America and southern Mexico.
    Through a case study related to the Asian summer monsoon (ASM), we verify that the intraseasonal changes of upper-level atmospheric conditions are preserved by significant connections within the global synchronization structure.
    Our work advances network-based clustering methodology for (i) decoding the spatiotemporal configuration of interdependence patterns of natural variability and for (ii) the intercomparison of these patterns, especially regarding their spatial distributions over different datasets.
\end{abstract}

\maketitle

\begin{quotation}
    {Precipitation variability of monsoons affects over two-thirds of the world’s population and regional monsoons have their own characteristics due to specific land–ocean and topographic conditions.~\cite{GM_Regional}
        In spite of being distributed in different continental regions, they are essentially driven and synchronized by the annual cycle of solar radiation.
        The connections between them are via the global divergent circulation characterized by a global-scale persistent overturning of the atmosphere varying with time.~\cite{GM,GM_Integration}
        An integration of these monsoons forms the concept of the global monsoon in terms of similar dynamics and behaviors.
        The synchronization in the context of the global monsoon is not limited to the tropical regions, but also extends to the subtropics, with the East Asian monsoon being a typical example.
        Therefore, identifying the synchronization structure on a global scale helps to understand the interaction {with mid-latitude regions}.
        A clustering workflow with higher robustness, {by combining consensus clustering and mutual correspondences,} is proposed for this purpose.}
\end{quotation}

\section{Introduction}\label{Introduction}
Regions of similar climatological properties are somewhat coherent in time and space.
Therefore, regional weather systems, formed within these areas, can exhibit synchronization behavior.
Synchronization is also observed in the climate system which spans thousands of kilometers.
This may be an indication of a teleconnection.~\cite{CN_1,CN_2}
Advances in the identification of these synchronized regions help to better understand the self-organized structure of the climate system and to further improve prediction skills.~\cite{Indices2,Indices3}

A complex-network-based clustering approach has proved to be useful in this respect, due to the climatological interpretation of the identified communities.~\cite{CD1}
It takes as input complex networks reconstructed from climate data.
Generally speaking, the reconstruction process treats grid points as nodes and establishes connections between nodes with a high correlation according to the corresponding timeseries.
Clustering such networks has been used for the discovery of spatial relationships in climate variables,~\cite{CD1,CD2,CD3} for the extraction of climate indices in a data-driven fashion,~\cite{Indices1,Indices2,Indices3} and for the intercomparison of performance of GCMs.~\cite{Model_Comparison}
{However, cluster analysis remains a challenging problem in this domain because of the availability of multiple datasets on each single climate variable.
    For example, each dataset of precipitation produces a network which further yields a community structure.
    How to extract representative and consistent communities when multiple community structures are obtained from different datasets on the same variable, is a key question.}

Regarding global-scale analysis,~\cite{CD1,CD2,CD3,CD4} previous studies based on network clustering put little focus on precipitation data, let alone EREs.
This is essentially due to the bias of precipitation estimation in reanalysis products.~\cite{CD1,ES_SAM_Precipitation}
However, it has recently been pointed out in the Sixth Assessment of the Intergovernmental Panel on Climate Change (IPCC), that precipitation extremes will be very likely to become more frequent in most locations.\footnote{https://www.ipcc.ch/assessment-report/ar6/}
Indeed, the investigation of synchronization behavior of extreme rainfall is of societal relevance due to the occurrences of natural hazards like floods or landslides.~\cite{ES_SAM_Precipitation, Reg_J_Rainfall}
A particular complex-network-based study on EREs~\cite{ES_First_NET} starts from the utilization of event synchronization (ES)~\cite{ES} as a nonlinear similarity measure for the reconstruction of the functional climate networks.
Using observational data, some studies emphasize how the network-based clustering decodes the spatial relationship on a regional scale.~\cite{Reg_Sw_Windspeed,Reg_India_Rainfall2,Reg_J_Rainfall,Reg_Australian_Monsoon}
A recent investigation reveals the global structure of synchronization of EREs based on high-resolution satellite data.~\cite{ES_Teleconnection}
But it relies on a given region of interest, and therefore it provides a partial view.
The focus of this study is therefore to reveal a comprehensive view of this global structure.

We address the above-mentioned issues through a systematic network-based clustering workflow.
Our aim is to unravel the synchronization of EREs around the globe, especially in terms of intraseasonal variability over different monsoon systems.
Specifically, the representation of climate data in the form of networks is implemented based on ES.~\cite{ES,ES_SAM_Precipitation,ES_Teleconnection}
The reconstructed functional climate networks are then viewed as input for a downstream cluster analysis.
Through this proposed clustering workflow, we identify two primary monsoon-related structures of distinct spatiotemporal characteristics during boreal summer.
Our work is the first combined application of consensus clustering~\cite{CSC} and mutual correspondences~\cite{Mutual_Cps} for climate extreme analysis.
This combination provides the identification of reliable interdependence patterns not only within each dataset, but also over different ones.

\section{Methodology}\label{Data and Methodology}

\begin{figure}[bt]
    \centering
    \includegraphics[width=0.45\textwidth]{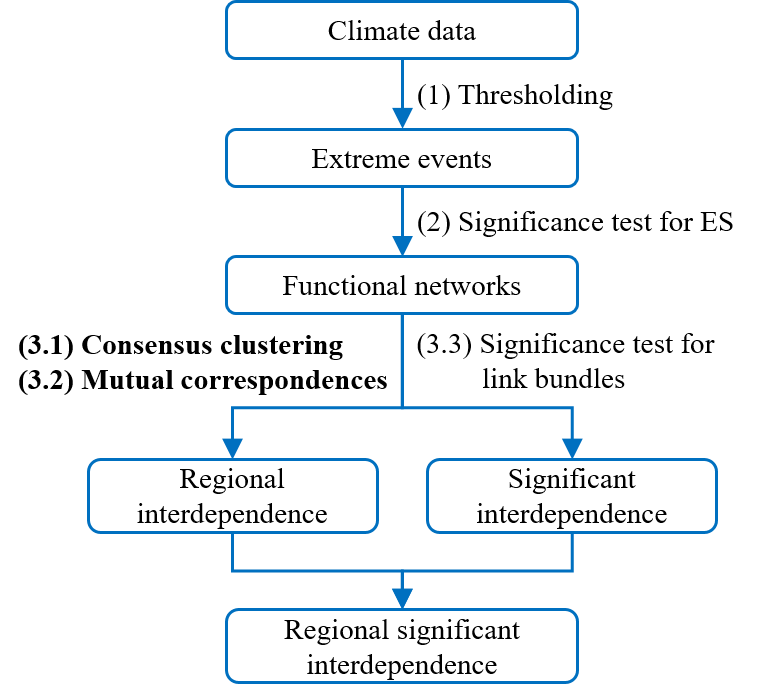}
    \caption{Schematic overview of the complex-network-based clustering workflow. Steps (1), (2), and (3.3) of the framework have been provided in Ref.~\onlinecite{ES_Teleconnection}. The new steps (3.1) and (3.2) are proposed in this paper.}
    \label{Work_Flow}
\end{figure}

We develop here a generic workflow {for the identification of the global synchronization structure of EREs,} as presented in Fig.~\ref{Work_Flow}.
The network reconstruction of functional climate networks, i.e., Steps~(1) and (2), is based on Ref.~\onlinecite{ES_Teleconnection}.
For the downstream analysis, our emphasis is on the proposed combined clustering workflow, in particular Steps~(3.1) and (3.2).
At the same time, we also include the correction for the multiple-comparison bias in Step (3.3) by using the technique introduced in Ref.~\onlinecite{ES_Teleconnection}.

\subsection{The Reconstruction of Functional Climate Networks}
\label{sub:reconstruction}
\paragraph*{(1) Definition of EREs based on Thresholding}~\\
For rainfall data, we adopt the $p = 90$th percentile of wet days (daily sum rainfall above 1mm), by the definition of extreme weather events from IPCC\footnote{https://www.ipcc.ch/report/ar5/wg1/}, as the threshold to indicate EREs.
Suppose that a pair of grid points $i$ and $j$ are randomly chosen from the entire set of grid cells.
For the two timeseries at $i$ and $j$, days with rainfall values higher than the given threshold are kept to indicate the occurrences of EREs.
For consecutive days with EREs, only the first day is preserved to represent an independent occurrence of events.~\cite{ES_Teleconnection}
We denote the set of grid points with at least three events as $V$.
The sets which incorporate the final event series, after data pre-processing at $i$ and $j$, are then defined as $e_i^p := \{t_{i, \alpha}\}$ and $e_j^p := \{t_{j, \beta}\}$ with $\alpha = 1,...,n_{i}, \beta = 1,...,n_{j}$, respectively.
For $i$, ${t_{i, \alpha}}$ denotes the time when the $\alpha$th event occurs and $n_{i}$ is the total number events.
Similarly, ${t_{j, \beta}}$ and $n_{j}$ are the corresponding variables for $j$.

\paragraph*{(2) ES and the Corresponding Significance Test}~\\
To quantify and determine the interactions between $i$ and $j$, ES considers all possible pairs of events ${t_{i, \alpha}}$ and ${t_{j, \beta}}$, and measures the closeness of each pair by imposing the condition that the absolute value of temporal delay $t_{i,j}^{\alpha, \beta} := |t_{i, \alpha} - t_{j, \beta}|$, must be smaller than a dynamical delay $\tau_{i,j}^{\alpha, \beta} := \frac{1}{2}\min\{t_{i,i}^{\alpha, \alpha - 1}, t_{i,i}^{\alpha, \alpha + 1}, t_{j,j}^{\beta, \beta - 1}, t_{j,j}^{\beta, \beta + 1}\}$.
Together with a maximum delay $\tau_{max}=10$ days, which is used to exclude the occurrences of unreasonably long delays, the $ES_{i,j}$ is hence defined between $e_i^p$ and $e_j^p$ as:~\cite{ES_Teleconnection}
\begin{equation}\label{ES_Equation}
    ES_{i,j} := |\{ (\alpha, \beta): t_{i,j}^{\alpha, \beta} < \tau_{i,j}^{\alpha, \beta} \wedge t_{i,j}^{\alpha, \beta} \leq \tau_{max} \}|,
\end{equation}
where $|\cdot |$ denotes the cardinality of a given set, i.e., the number of synchronized event pairs.

The placement of network links between each pair of grid points is determined statistically.
Taking $i$ and $j$ as an example, by randomly distributing the same number of events 2000 times within the June-July-August (JJA) season and computing the corresponding ES values for the 2000 pairs of surrogate event series, a null-model distribution is obtained.~\cite{ES_Teleconnection}
We take the $99.5$th percentile of the distribution as the threshold.
A significant connection with $P < 0.005$ is determined if $ES_{i,j}$ is above the threshold.
Since the significance thresholds are only associated with the number of events, we manage this process independently for all possible pairs of event numbers.

In many scenarios, links of a network can be attached with features such as weight and direction.
For the reconstructed functional climate networks, each significant link has an ES value as a weight to indicate the link strength.
This is an important factor in Step (3.1) (see Sec.~\ref{The network-based clustering workflow}) to determine the affinities of different nodes.
Regarding link direction, it can also be incorporated into the calculation of ES and the corresponding test of significance.
For example, the synchronization from $i$ to $j$ is defined as:
\begin{equation}\label{ESij_Equation}
    ES_{i \to j} := |\{ (\alpha, \beta): t_{i \to j}^{\alpha, \beta} < \tau_{i,j}^{\alpha, \beta} \wedge t_{i \to j}^{\alpha, \beta} \leq \tau_{max} \}|,
\end{equation}
where $t_{i \to j}^{\alpha, \beta} := t_{j, \beta} - t_{i, \alpha}$ and $t_{i \to j}^{\alpha, \beta} > 0$ indicate the occurrence of an event at $i$ with a subsequent one at $j$.
The significance test for the directed case follows the same procedure as the one for the undirected one.
We take the undirected weighted network as the default setting over the following main text.
The directed approach is only used when identifying days of high synchronization for atmospheric condition analysis in Sec.~\ref{The network-based clustering workflow}.

\subsection{The Network-based Clustering Workflow}
\label{The network-based clustering workflow}
\paragraph*{(3.1) Consensus Clustering}~\\
The significant link bundles given by Step (3.3)~\cite{ES_Teleconnection} have limited capability to capture regions following similar climatological behavior around the globe.
This is essentially due to the limited link distribution based on a selected region (see~\hyperref[ESLB_1998_To_2019_JJA_900_d10_P995_LBP999_X4_Exp_Tele]{Fig.}~\ref{ESLB_1998_To_2019_JJA_900_d10_P995_LBP999_X4_Exp_Tele} in~\cref{THE MOTIVATION FOR THE NETWORK-BASED WORKFLOW} as an example).
Therefore, we adopt here complex-network-based clustering, namely, community detection, to fill this gap, by taking into account the link distribution of a whole reconstructed functional network.
{However, most community detection methods are not deterministic, indicating that partitions delivered by them have certain fluctuations, e.\ g.\ due to randomness.
One way to manage this problem is to use consensus clustering.~\cite{CSC}
Using the consensus of a combination of partitions provides a stable result out of a set of candidates.
In the context of climate science, to the best of our knowledge, the use of consensus clustering
represents a novelty; using a stable partition derived this way yields an algorithmically reliable and climatologically interpretable community structure for a given dataset.

Given the reconstructed functional climate network $G$ with $|V|$ nodes and a community detection method $A$, below is the detailed procedure of consensus clustering:

\begin{enumerate}[({A}1)]
    \item Apply $A$ on $G$ for 1000 times, yielding 1000 partitions.
    \item Rank the 1000 partitions based on modularity~\cite{Modularity} and select the first $n_p$ partitions of the highest modularity.
    \item Compute the consensus matrix $D$: $D_{ij}$ is the number of partitions where nodes $i$ and $j$ belong to the same community, divided by $n_p$.
    \item All entries of $D$ below a chosen threshold $\theta$ are set to zero.
    \item Apply $A$ on $D$ for $n_p$ times, yielding $n_p$ partitions.
    \item If the partitions in {(A5)} are all equal, that is, $D$ is block-diagonal, then stop; otherwise, go back to {(A3)}.
\end{enumerate}
Note that $A$ serves as a basic method for the entire procedure and we use modularity to estimate the quality/strength of a partition, due to the lack of climate-related benchmarks.
Modularity compares the fraction of intra-community links with its expected number in a null model,
which is a network with the same degree sequence but links placed at random.~\cite{Modularity}
Intuitively speaking, networks with high modularity have dense intra-community connections,
but are sparsely connected between different communities.}

{{In (A1)}, we use the parallel Louvain method (PLM)~\cite{PLM} as the basic method as algorithm $A$.
We choose PLM because
    {(a)} it is flexible enough to control the granularity/size of communities with only one parameter $\gamma$,
{(b)} its speed allows to handle large-scale networks, and
    {(c)} it is based on the well-known Louvain method.~\cite{LM}
The Louvain method is a greedy local search algorithm with a bottom-up multilevel approach.
Specifically, the local search phase moves each node to the neighbor community where this move yields the largest increase in modularity (if any).
After this process has converged, each community is contracted to a super-node.
The super-nodes are linked together by weighted edges [and self-loops] whose weights are set according to the inter-community [intra-community] edge weights in the larger
graph.
The smaller network derived this way is treated as new input for the next iteration of
local search and coarsening (until no further community changes are observed).
Among others, the main change of PLM compared to plain Louvain is to use parallelism.
PLM is part of the network analysis toolkit NetworKit,~\cite{Networkit} which allows the integration of PLM into more complex workflows.}
We leave the default setting $\gamma=1$, since this already yields an appropriate community resolution.
The 1000 partitions are particularly generated in search of a broad range of possible solutions, since the inherent uncertainty due to non-determinism due to thread parallelism can easily change solutions, especially for those algorithms relying on stochastic search strategies (see~\hyperref[CD_Random_Noise]{Fig.}~\ref{CD_Random_Noise} in~\cref{THE MOTIVATION FOR THE NETWORK-BASED WORKFLOW} as an example).~\cite{CSC}
{In (A2)}, we choose $n_p \in \{25, 50, 100\}$ candidates of the highest modularity as input for reaching a consensus.
    {In (A4)}, the optimal $\theta$ is situation-dependent.~\cite{CSCthreshold}
Because this threshold parameter also implies the probability of two nodes being settled into the same community, we vary it in our work and let $\theta \in \{0.5, 0.6, 0.7, 0.8, 0.9\}$.
    {In (A6)}, considering multiple partitions to be obtained under different values of $n_p$ and $\theta$, the final decision is made based on the modularity score again.

\paragraph*{(3.2) Mutual Correspondences}~\\
Comparing different communities based on single similarity values does neither provide information on how they differ nor any spatial information on the robustness of the identified interdependence patterns of EREs.
That is why we employ mutual correspondences,~\cite{Mutual_Cps} which provide insights on how
different partitions agree with each other on a meta-community level (i.\ e., a group of communities).
Suppose that there are two stable partitions $C=\{c_1, \dots, c_{|C|}\}$ and $C^{'}=\{c^{'}_1, \dots, c^{'}_{|C^{'}|}\}$ obtained, for example from two datasets.
Given $\emptyset \neq S\subsetneq C$ and $\emptyset \neq S^{'}\subsetneq C^{'}$, a correspondence $(S, S^{'})$ is mutual when the following conditions are met:
a. $|\cup_{i=1}^{|S|}S_{i} \cap s^{'}| > \frac{1}{2}|s^{'}|$ for any $s^{'} \in S^{'}$;
b. $|\cup_{i=1}^{|S|}S_{i} \cap s^{'}| \leq \frac{1}{2}|s^{'}|$ for any $s^{'} \in C^{'} \setminus S^{'}$;
c. $|\cup_{i=1}^{|S^{'}|}S^{'}_{i} \cap s| > \frac{1}{2}|s|$ for any $s \in S$;
d. $|\cup_{i=1}^{|S^{'}|}S^{'}_{i} \cap s| \leq \frac{1}{2}|s|$ for any $s \in C \setminus S$.
The four conditions above guarantee that any community $s$ from $S$ preserves more than 50\% similarity to the counterpart $s^{'}$ in $S^{'}$, and vice versa.
{We explain below the detailed procedure in finding such a mutual correspondence:
\begin{enumerate}[({B}1)]
    \item Initialize $S$ with $S_{t}$ and $\emptyset \neq S_{t}\subsetneq C$. Only for the first iteration, $S_{t}$ is chosen based on domain knowledge such as well-known teleconnection patterns and is preserved by $S_{init} = S_{t}$ for the final check {in (B5)}; otherwise, {$S$ equals $S_{t}$ which is found in (B3).}
    \item Each element $c^{'} \in C^{'}$ satisfying $|\cup_{i=1}^{|S|}S_{i} \cap c^{'}| > \frac{1}{2}|c^{'}|$, where $S=S_{t}$ as {in (B1)}, is placed into $S^{'}_{t}$, with the correspondence direction of $S_{t} \rightarrow S^{'}_{t}$.
    \item Each element $c \in C$ satisfying $|\cup_{i=1}^{|S^{'}|}S^{'}_{i} \cap c| > \frac{1}{2}|c|$, where $S^{'}$ equals $S^{'}_{t}$ found {in (B2)}, is placed into $S_{t}$, with the reversed correspondence direction of $S^{'}_{t} \rightarrow S_{t}$.
    \item If $(S_{t} \cup S^{'}_{t})_{S_{t} \rightarrow S^{'}_{t}}$ given {in (B2)} equals $(S^{'}_{t} \cup S_{t})_{S^{'}_{t} \rightarrow S_{t}}$ given {in (B3)}, then stop; otherwise, go back to {(B1)} for the next iteration.
    \item If the updated $S_t$ and $S^{'}_t$ still satisfy $\emptyset \neq S_t\subsetneq C$, $\emptyset \neq S^{'}_t\subsetneq C^{'}$, and $S_{init} \subseteq S_t$ when the iteration process stops in {(B4)}, then a mutual correspondence $(S=S_t, S^{'}=S^{'}_{t})$ is established.
\end{enumerate}
It is also possible to start by initializing $S^{'}$ {in (B1)}.
The resulting correspondence relationship {is $(S^{'}=S^{'}_{t}, S=S_t)$, which is equal to $(S=S_t, S^{'}=S^{'}_{t})$ obtained by initializing $S$ in (B1).}}

Together with consensus clustering, mutual correspondences lead to a robust top-down clustering workflow. Both combined allow a reliable identification of synchronization structure in the context of global monsoon.
A mutual correspondence can be one-to-many or many-to-one to indicate the split or combination of rainfall patterns over different datasets.
For a many-to-many correspondence, it is reasonable to only consider neighborhood communities due to spatial coherence.
Finally, we estimate the meta-community level similarity based on $RCR(S, S^{'}) = \frac{|S \cap S^{'} |}{|S \cup S^{'} |}$.

\paragraph*{Identification of synchronized days for meta-communities.}
After identifying meta-communities of good mutual correspondence, we further determine synchronized days for them.
Assume that $(t_{i, \alpha}, t_{j, \beta})$ is one pair of synchronized events when $ES_{i,j}$ is significant.
It is therefore plausible to take $\min\{t_{i, \alpha}, t_{j, \beta}\}$ as the day to indicate the occurrence of a synchronization.
By repeating this process for all synchronized pairs between $i$ and $j$ and for all other significant links within a meta-community, we obtain the distribution of these days over the JJA season.
Such a distribution indicates the temporal characteristic of a given meta-community.

\paragraph*{(3.3) Significance Test for Link Bundles}~\\
During the (re)construction process of functional climate networks, each link is determined
statistically by multiple comparisons in Step~(2).~\cite{ES_Teleconnection}
A large number of comparison can lead to some links being preserved by chance.
Therefore, we need to correct this problem afterwards and to complement the proposed clustering workflow.
The basic idea is to identify some regions of higher link density formed by physical mechanisms.
Specifically, given a region and all links coming from this region, a spherical Gaussian kernel density estimate (KDE) provides the link density estimation for each grid point regarding the distribution of the original regional link configuration.
By randomly distributing those links 1000 times to grid points with at least three events, a corresponding null-model distribution based on the same KDE is obtained for each grid point.
The thresholds are then taken in the same way as in Step (2).
We also use the $99.9$th percentile to determine statistically significant link bundles ($P < 0.001$).



The combination of Steps (3.1), (3.2), and (3.3) gives the regional significant synchronization structure of EREs.
Specifically, the regional interdependence by Steps (3.1) and (3.2) takes the form of meta-communities.
For a specified region {`A'} within a meta-community {$S$ ($S^{'}$)}, we determine all other regions that are significantly synchronized with `A' based on Step (3.3).
Some of these regions fall into {$S$ ($S^{'}$)}, forming the final significantly synchronized regional structure {associated with `A'}.
Such a structure could indicate rainfall teleconnections, which can further be verified by atmospheric conditions.

\subsection{Atmospheric Condition Analysis}\label{Atmospheric condition analysis}
{This part is the verification for the identified synchronization structure.
    Therefore, it is not indicated in the workflow (see Fig.~\ref{Work_Flow}).}
Specifically, to investigate composite anomalies, we first identify days of high synchronization.
Suppose that we need to find out the specific days when two regions `A' and `B' are highly synchronized.
We first determine significant links between them, with direction from `A' (`B') to `B' (`A'), under a significance level of $P<0.005$.
Then, for each time step, we identify the number of associated EREs occurring in `B' (`A') within $\tau_{max}=10$ days, when there are events in `A' (`B'), based on these links.
All these event numbers form a timeseries from `A' (`B') to `B' (`A'), with the same time steps as datasets.
They are then both filtered by a Butterworth low-pass filter with a cutoff frequency of 10 days.
We take those time steps as days of high synchronization, where there are local maxima above the $90$th percentile of JJA season in the filtered timeseries.
With these days, we calculate the composite anomalies regarding the JJA climatology.

\section{Results}\label{Result}
\subsection{Data}
For the reconstruction of functional climate networks regarding EREs, we choose observational data due to the bias of precipitation estimation in reanalysis products.~\cite{CD1}
Two satellite-derived datasets, i.e., TRMM 3B42 V7~\cite{TRMM} and GPM IMERG~\cite{GPM} from NASA, are analyzed here.
We utilize the daily rainfall sums within the JJA season from 2000 to 2019 with the spatial resolution of 1$^{\circ}$ $\times$ 1$^{\circ}$.
For a comparative analysis to TRMM, the GPM data are cropped with a boundary from 50$^{\circ}$S to 50$^{\circ}$N, named as B-GPM.

For the understanding of climatological behavior behind the obtained rainfall patterns, we choose horizontal and vertical wind fields from ECMWF (ERA5),~\cite{ERA5} to get the corresponding atmospheric conditions.
The spatial (i.e., 1$^{\circ}$ $\times$ 1$^{\circ}$) and temporal (i.e., daily estimations ranging from 2000 to 2019) resolutions are consistent with TRMM.

\begin{figure*}[t]
    \centering
    \includegraphics[width=.95\textwidth]{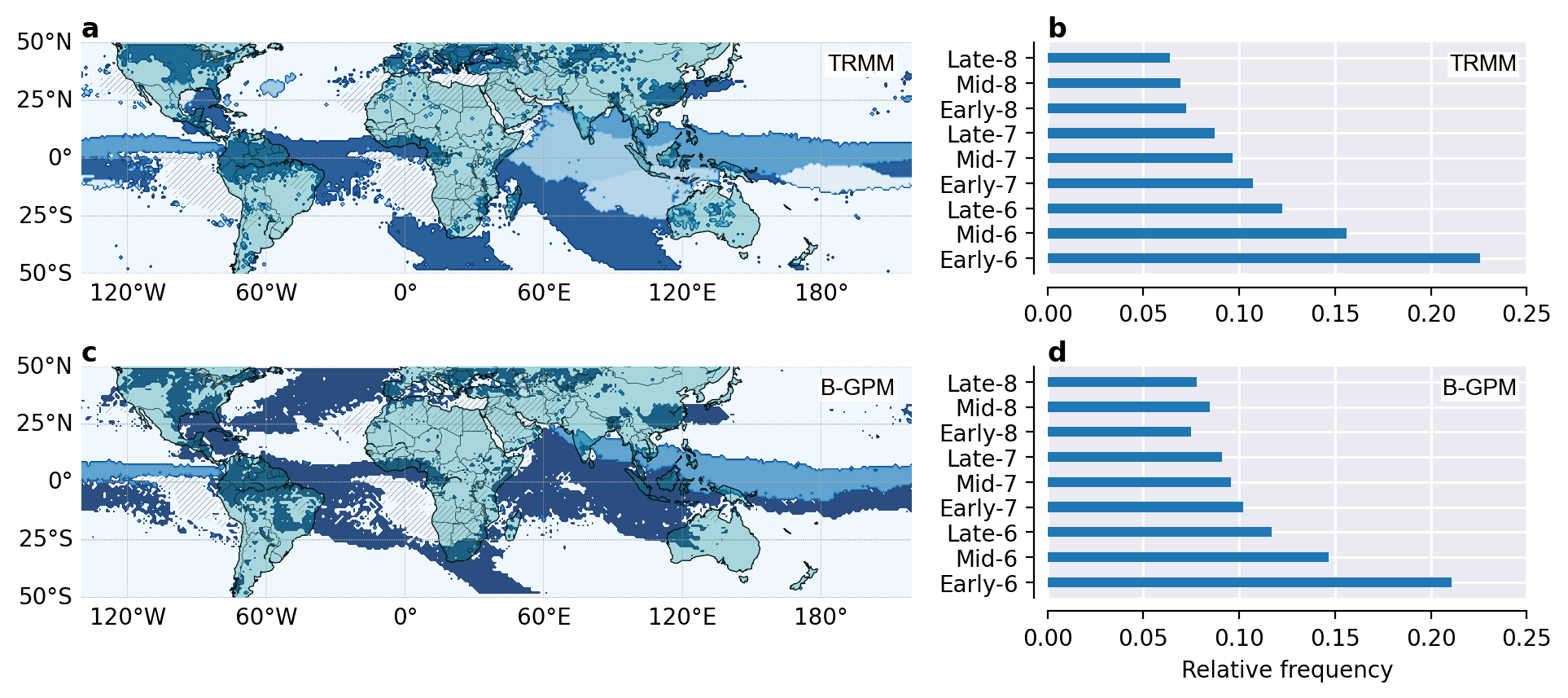}
    \caption{Communities of mutual correspondence with a similarity of $RCR($TRMM=\{2, 6, 9, 10, 13\}, B-GPM=\{1, 5\}$)=0.54$. (a) Communities 2, 6, 9, 10 and 13 from~\hyperref[TRMM_GPMAST_LR_d10_P995]{Fig.}~\ref{TRMM_GPMAST_LR_d10_P995}\hyperref[TRMM_GPMAST_LR_d10_P995]{(a)} in~\cref{COMMUNITY STRUCTURE UNDER THE DELAY OF 10 DAYS}. (c) Communities 1 and 5 from ~\hyperref[TRMM_GPMAST_LR_d10_P995]{Fig.}~\ref{TRMM_GPMAST_LR_d10_P995}~\hyperref[TRMM_GPMAST_LR_d10_P995]{(b)} in~\cref{COMMUNITY STRUCTURE UNDER THE DELAY OF 10 DAYS}. (b) and (d): Frequency of synchronized days over JJA, within meta-communities in (a) and (c), respectively (see Sec.~\ref{The network-based clustering workflow} `Identification of synchronized days for meta-communities'). Grid points with less than three events are excluded when extracting EREs and visualized as hatched areas here.
    }
    \label{TRMM_GPMAST_CPDS_ES_2000_To_2019_JJA_900_d10_P995_1X1_T2691013G15}
\end{figure*}

\begin{figure*}[!htbp]
    \centering
    \includegraphics[width=.95\textwidth]{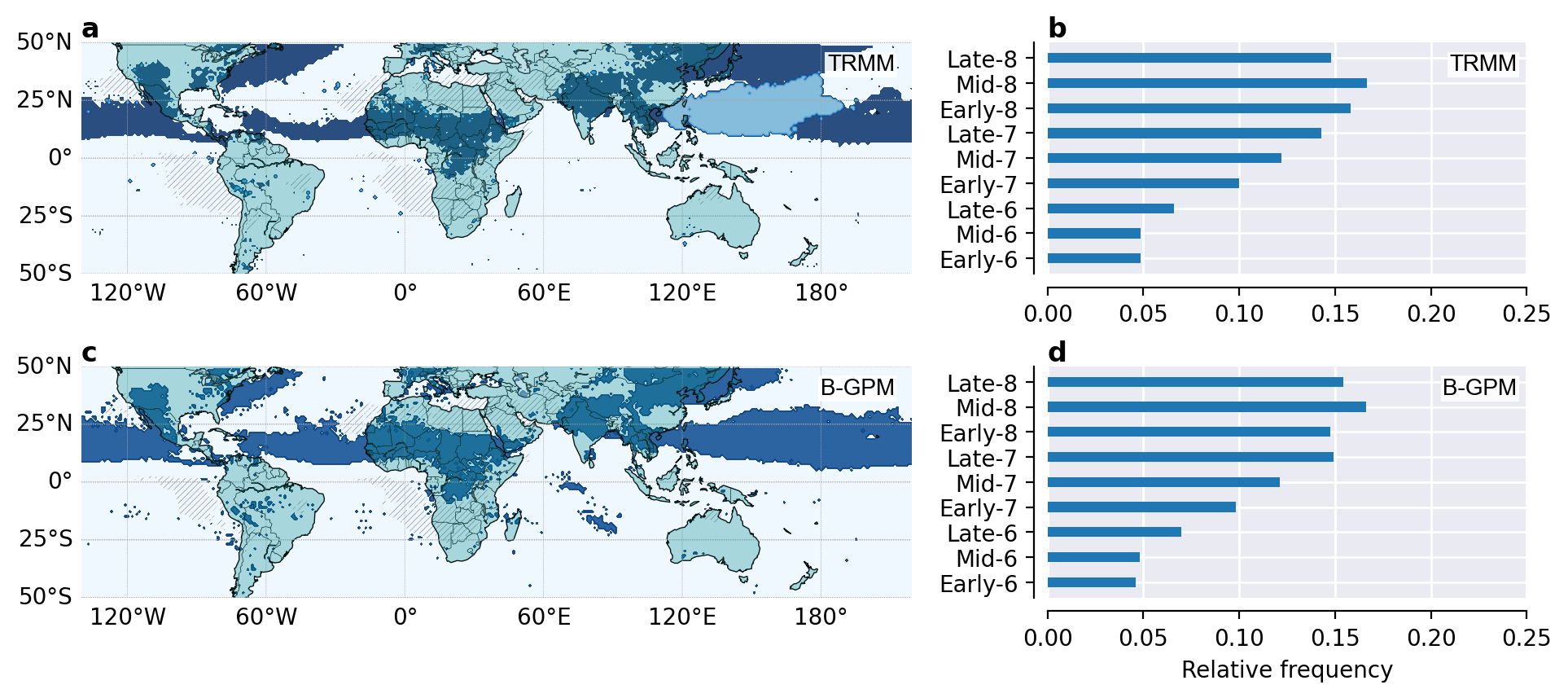}
    \caption{Same as~\hyperref[TRMM_GPMAST_CPDS_ES_2000_To_2019_JJA_900_d10_P995_1X1_T2691013G15]{Fig.}~\ref{TRMM_GPMAST_CPDS_ES_2000_To_2019_JJA_900_d10_P995_1X1_T2691013G15}, but for a different mutual correspondence with a similarity of $RCR($TRMM=\{1, 8\}, B-GPM=\{2\}$)=0.63$ from~\hyperref[TRMM_GPMAST_LR_d10_P995]{Fig.}~\ref{TRMM_GPMAST_LR_d10_P995} in~\cref{COMMUNITY STRUCTURE UNDER THE DELAY OF 10 DAYS}.
    }
    \label{TRMM_GPMAST_CPDS_ES_2000_To_2019_JJA_900_d10_P995_1X1_T18G2}
\end{figure*}

\subsection{Regional interdependence of EREs}
{The global monsoon represents the integration of all regional monsoons.
Different regional monsoons are synchronized, since they are fundamentally driven by the annual variation in solar radiation.~\cite{GM_Integration}
In the context of the global monsoon, the following two points need to be further explored:
{(i)} The synchronization structure with higher latitudes;
{(ii)} The spatiotemporal characteristics of intraseasonal variability.
We apply the proposed clustering workflow (see Fig.~\ref{Work_Flow}) to the global precipitation datasets and give appropriate interpretations of the results.}

Two main monsoon-related meta-communities are identified during the JJA season, as shown in~\hyperref[TRMM_GPMAST_CPDS_ES_2000_To_2019_JJA_900_d10_P995_1X1_T2691013G15]{Figs.}~\ref{TRMM_GPMAST_CPDS_ES_2000_To_2019_JJA_900_d10_P995_1X1_T2691013G15} and~\ref{TRMM_GPMAST_CPDS_ES_2000_To_2019_JJA_900_d10_P995_1X1_T18G2}.
An overview of the entire community structure is presented in~\hyperref[TRMM_GPMAST_LR_d10_P995]{Fig.}~\ref{TRMM_GPMAST_LR_d10_P995} in~\cref{COMMUNITY STRUCTURE UNDER THE DELAY OF 10 DAYS}.
{The meta-communities describe a global view of synchronized EREs, with a synchronization structure extending to mid-latitude regions {(to point (i))} and distinct intraseasonal changes on the frequency of synchronization {(to point (ii)).}}
Temporal distributions indicate two recognizable periods of synchronization: from early June to mid-July (see~\hyperref[TRMM_GPMAST_CPDS_ES_2000_To_2019_JJA_900_d10_P995_1X1_T2691013G15]{Figs.}~\ref{TRMM_GPMAST_CPDS_ES_2000_To_2019_JJA_900_d10_P995_1X1_T2691013G15}\hyperref[TRMM_GPMAST_CPDS_ES_2000_To_2019_JJA_900_d10_P995_1X1_T2691013G15]{(b)} and~\ref{TRMM_GPMAST_CPDS_ES_2000_To_2019_JJA_900_d10_P995_1X1_T2691013G15}\hyperref[TRMM_GPMAST_CPDS_ES_2000_To_2019_JJA_900_d10_P995_1X1_T2691013G15]{(d)}), and from mid-July to late-August (see~\hyperref[TRMM_GPMAST_CPDS_ES_2000_To_2019_JJA_900_d10_P995_1X1_T18G2]{Figs.}~\ref{TRMM_GPMAST_CPDS_ES_2000_To_2019_JJA_900_d10_P995_1X1_T18G2}\hyperref[TRMM_GPMAST_CPDS_ES_2000_To_2019_JJA_900_d10_P995_1X1_T18G2]{(b)} and~\ref{TRMM_GPMAST_CPDS_ES_2000_To_2019_JJA_900_d10_P995_1X1_T18G2}\hyperref[TRMM_GPMAST_CPDS_ES_2000_To_2019_JJA_900_d10_P995_1X1_T18G2]{(d)}).
    {The former peaks in early-June with a sharply decreasing trend thereafter, while the latter shows a gradual increase with a peak in mid-August.
        Since the temporal distributions are obtained based on the timing of all synchronous EREs within the identified meta-communities, it can be concluded that those teleconnection patterns carried by significant connections within meta-communities are also experiencing a similar temporal variability.
        We verify this with the ASM-related case study in Sec.~\ref{Regional significant interdependence of EREs: a case study related to ASM}.}
Regarding the corresponding spatial distributions, they decode the compositions of regional EREs, including different monsoon systems.
They are robust under different choices of maximum delays (see~\hyperref[TRMM_GPMAST_CPDS_ES_2000_To_2019_JJA_900_d3_P995_1X1_T269G2]{Figs.}~\ref{TRMM_GPMAST_CPDS_ES_2000_To_2019_JJA_900_d3_P995_1X1_T269G2}-\ref{TRMM_GPMAST_CPDS_ES_2000_To_2019_JJA_900_d30_P995_1X1_T26T29} in~\cref{THE ROBUSTNESS TEST ON DIFFERENT DELAYS}) and significance levels (see~\hyperref[TRMM_GPMAST_CPDS_ES_2000_To_2019_JJA_900_d10_P994_1X1_T2691013G15]{Figs.}~\ref{TRMM_GPMAST_CPDS_ES_2000_To_2019_JJA_900_d10_P994_1X1_T2691013G15}-\ref{TRMM_GPMAST_CPDS_ES_2000_To_2019_JJA_900_d10_P996_1X1_T110G2} in~\cref{THE ROBUSTNESS TEST ON DIFFERENT SIGNIFICANCE LEVELS}).
    {Based on regional monsoons over Asia, West Africa, and Southwest United States in boreal summer, we now provide detailed evidence corroborating the spatiotemporal characteristics of intraseasonal variability found in~\hyperref[TRMM_GPMAST_CPDS_ES_2000_To_2019_JJA_900_d10_P995_1X1_T2691013G15]{Figs.}~\ref{TRMM_GPMAST_CPDS_ES_2000_To_2019_JJA_900_d10_P995_1X1_T2691013G15} and~\ref{TRMM_GPMAST_CPDS_ES_2000_To_2019_JJA_900_d10_P995_1X1_T18G2}.}

Our results are in line with the known teleconnection pattern inside of ASM.
The Meiyu/Baiu rainband,~\cite{EASM_Overview} as in~\hyperref[TRMM_GPMAST_CPDS_ES_2000_To_2019_JJA_900_d10_P995_1X1_T2691013G15]{Figs.}~\ref{TRMM_GPMAST_CPDS_ES_2000_To_2019_JJA_900_d10_P995_1X1_T2691013G15}\hyperref[TRMM_GPMAST_CPDS_ES_2000_To_2019_JJA_900_d10_P995_1X1_T2691013G15]{(a)} and~\ref{TRMM_GPMAST_CPDS_ES_2000_To_2019_JJA_900_d10_P995_1X1_T2691013G15}\hyperref[TRMM_GPMAST_CPDS_ES_2000_To_2019_JJA_900_d10_P995_1X1_T2691013G15]{(c)}, is automatically identified by using the proposed clustering workflow.
The coexistence of this noticeable region, the Bay of Bengal, and the western coast of India in the same meta-community, supports the synchronized initiation (i.e., the `south' teleconnection) of the Indian rainy season and Meiyu/Baiu.~\cite{EASM_Overview,Southern_Teleconnection}
\hyperref[TRMM_GPMAST_CPDS_ES_2000_To_2019_JJA_900_d10_P995_1X1_T18G2]{Figures}~\ref{TRMM_GPMAST_CPDS_ES_2000_To_2019_JJA_900_d10_P995_1X1_T18G2}\hyperref[TRMM_GPMAST_CPDS_ES_2000_To_2019_JJA_900_d10_P995_1X1_T18G2]{(a)} and~\ref{TRMM_GPMAST_CPDS_ES_2000_To_2019_JJA_900_d10_P995_1X1_T18G2}\hyperref[TRMM_GPMAST_CPDS_ES_2000_To_2019_JJA_900_d10_P995_1X1_T18G2]{(c)} preserve the `north' teleconnection between India and North China, which is built-up after a rapid northward jump of the rain belt.~\cite{Northern_Teleconnection}
For adjacent oceanic areas, the connection between Indonesian rainfall with Indian and Pacific Oceans~\cite{ENSO_IOD_Indonesia} is partly revealed by the coexistence of these regions in~\hyperref[TRMM_GPMAST_CPDS_ES_2000_To_2019_JJA_900_d10_P995_1X1_T2691013G15]{Fig.}~\ref{TRMM_GPMAST_CPDS_ES_2000_To_2019_JJA_900_d10_P995_1X1_T2691013G15}.
In~\hyperref[TRMM_GPMAST_CPDS_ES_2000_To_2019_JJA_900_d10_P995_1X1_T18G2]{Fig.}~\ref{TRMM_GPMAST_CPDS_ES_2000_To_2019_JJA_900_d10_P995_1X1_T18G2}, the northwestward propagation of cyclones in a high phase of the Pacific-Japan pattern establishes the connection of EREs between the western North Pacific tropics and the northern South China Sea; during a low phase, cyclones follow a recurved propagation route to a higher latitude, close to the east of Japan.~\cite{PJ}
These cyclones form the synchronization structure relating to the global monsoon, especially over oceanic areas.

\hyperref[TRMM_GPMAST_CPDS_ES_2000_To_2019_JJA_900_d10_P995_1X1_T2691013G15]{Figs.}~\ref{TRMM_GPMAST_CPDS_ES_2000_To_2019_JJA_900_d10_P995_1X1_T2691013G15} and~\ref{TRMM_GPMAST_CPDS_ES_2000_To_2019_JJA_900_d10_P995_1X1_T18G2} capture the abrupt intraseasonal change in West Africa, i.\ e., from the oceanic regime (5$^{\circ}$ North of equator, see~\hyperref[TRMM_GPMAST_CPDS_ES_2000_To_2019_JJA_900_d10_P995_1X1_T2691013G15]{Figs.}~\ref{TRMM_GPMAST_CPDS_ES_2000_To_2019_JJA_900_d10_P995_1X1_T2691013G15}\hyperref[TRMM_GPMAST_CPDS_ES_2000_To_2019_JJA_900_d10_P995_1X1_T2691013G15]{(a)} and~\ref{TRMM_GPMAST_CPDS_ES_2000_To_2019_JJA_900_d10_P995_1X1_T2691013G15}\hyperref[TRMM_GPMAST_CPDS_ES_2000_To_2019_JJA_900_d10_P995_1X1_T2691013G15]{(c)}) to the continental regime (10$^{\circ}$ North of equator, see~\hyperref[TRMM_GPMAST_CPDS_ES_2000_To_2019_JJA_900_d10_P995_1X1_T18G2]{Figs.}~\ref{TRMM_GPMAST_CPDS_ES_2000_To_2019_JJA_900_d10_P995_1X1_T18G2}\hyperref[TRMM_GPMAST_CPDS_ES_2000_To_2019_JJA_900_d10_P995_1X1_T18G2]{(a)} and~\ref{TRMM_GPMAST_CPDS_ES_2000_To_2019_JJA_900_d10_P995_1X1_T18G2}\hyperref[TRMM_GPMAST_CPDS_ES_2000_To_2019_JJA_900_d10_P995_1X1_T18G2]{(c)}).~\cite{AM_Phase1}
Between Indian and African monsoons, there exists a connection provided by westward-propagating Rossby waves from the Pacific warm pool in boreal summer.~\cite{ISM_AM_Teleconnection1}
They help modulate the easterly wave activity and moisture transport in West Africa.~\cite{ISM_AM_Teleconnection2}
Meta-communities in~\hyperref[TRMM_GPMAST_CPDS_ES_2000_To_2019_JJA_900_d10_P995_1X1_T2691013G15]{Figs.}~\ref{TRMM_GPMAST_CPDS_ES_2000_To_2019_JJA_900_d10_P995_1X1_T2691013G15}\hyperref[TRMM_GPMAST_CPDS_ES_2000_To_2019_JJA_900_d10_P995_1X1_T2691013G15]{(a)} and~\ref{TRMM_GPMAST_CPDS_ES_2000_To_2019_JJA_900_d10_P995_1X1_T2691013G15}\hyperref[TRMM_GPMAST_CPDS_ES_2000_To_2019_JJA_900_d10_P995_1X1_T2691013G15]{(c)}, covering the Gulf of Guinea with extension to tropical South America, Caribbean Sea, and the eastern tropical Pacific, manifest the underlying connection via easterly wave disturbances over the equatorial Atlantic basin.~\cite{EWD1,EWD2}
These disturbances contribute to a large proportion of rainfall in these tropical regions.
In~\hyperref[TRMM_GPMAST_CPDS_ES_2000_To_2019_JJA_900_d10_P995_1X1_T18G2]{Figs.}~\ref{TRMM_GPMAST_CPDS_ES_2000_To_2019_JJA_900_d10_P995_1X1_T18G2}\hyperref[TRMM_GPMAST_CPDS_ES_2000_To_2019_JJA_900_d10_P995_1X1_T18G2]{(a)} and~\ref{TRMM_GPMAST_CPDS_ES_2000_To_2019_JJA_900_d10_P995_1X1_T18G2}\hyperref[TRMM_GPMAST_CPDS_ES_2000_To_2019_JJA_900_d10_P995_1X1_T18G2]{(c)}, meta-communities over West Africa experience a northward shift which is also known as `monsoon jump'.\cite{AM_Jump,AM_Overview}
Along with this, easterly waves emanating from West Africa are enhanced in August over the tropical North Atlantic.~\cite{AEW_Cyclones1}
They are the primary precursors of tropical cyclones which propagate toward the Caribbean Sea and North America.~\cite{ES_Teleconnection,AEW_Cyclones2,AEW_Cyclones3,AEW_Cyclones4}
It has also been proposed that the easterly waves from West Africa have a weak correlation with tropical cyclogenesis in the eastern North Pacific.~\cite{AEW_Cyclones5_Weak}
Such a correlation provides a possible explanation for the synchronized EREs between them (see~\hyperref[TRMM_GPMAST_CPDS_ES_2000_To_2019_JJA_900_d10_P995_1X1_T18G2]{Figs.}~\ref{TRMM_GPMAST_CPDS_ES_2000_To_2019_JJA_900_d10_P995_1X1_T18G2}\hyperref[TRMM_GPMAST_CPDS_ES_2000_To_2019_JJA_900_d10_P995_1X1_T18G2]{(a)} and~\ref{TRMM_GPMAST_CPDS_ES_2000_To_2019_JJA_900_d10_P995_1X1_T18G2}\hyperref[TRMM_GPMAST_CPDS_ES_2000_To_2019_JJA_900_d10_P995_1X1_T18G2]{(c)}).

For surrounding regions related to the Southwest United States monsoon, our results catch the mid-summer drought from July to August, especially over Central America, southern Mexico, and part of the Caribbean Sea (see~\hyperref[TRMM_GPMAST_CPDS_ES_2000_To_2019_JJA_900_d10_P995_1X1_T18G2]{Figs.}~\ref{TRMM_GPMAST_CPDS_ES_2000_To_2019_JJA_900_d10_P995_1X1_T18G2}\hyperref[TRMM_GPMAST_CPDS_ES_2000_To_2019_JJA_900_d10_P995_1X1_T18G2]{(a)} and~\ref{TRMM_GPMAST_CPDS_ES_2000_To_2019_JJA_900_d10_P995_1X1_T18G2}\hyperref[TRMM_GPMAST_CPDS_ES_2000_To_2019_JJA_900_d10_P995_1X1_T18G2]{(c)}).~\cite{NAM_Drought1,NAM_Drought2}
In June, the intensified Caribbean and Great Plains low-level jets together contribute to the moisture supply to rainfall in parts of North America.~\cite{NAM_Sync1}
This leads to the synchronized EREs along the way from the Caribbean Sea to the central United States (see~\hyperref[TRMM_GPMAST_CPDS_ES_2000_To_2019_JJA_900_d10_P995_1X1_T2691013G15]{Figs.}~\ref{TRMM_GPMAST_CPDS_ES_2000_To_2019_JJA_900_d10_P995_1X1_T2691013G15}\hyperref[TRMM_GPMAST_CPDS_ES_2000_To_2019_JJA_900_d10_P995_1X1_T2691013G15]{(a)} and~\ref{TRMM_GPMAST_CPDS_ES_2000_To_2019_JJA_900_d10_P995_1X1_T2691013G15}\hyperref[TRMM_GPMAST_CPDS_ES_2000_To_2019_JJA_900_d10_P995_1X1_T2691013G15]{(c)}).

Apart from the above-mentioned synchronization related to monsoon, it is also observed that the meta-communities in~\hyperref[TRMM_GPMAST_CPDS_ES_2000_To_2019_JJA_900_d10_P995_1X1_T2691013G15]{Figs.}~\ref{TRMM_GPMAST_CPDS_ES_2000_To_2019_JJA_900_d10_P995_1X1_T2691013G15} and~\ref{TRMM_GPMAST_CPDS_ES_2000_To_2019_JJA_900_d10_P995_1X1_T18G2} span the mid-latitude belt.
The connections between them are mainly via upper-level teleconnection patterns like the `Silk Road teleconnection'~\cite{SR} and the `circumglobal teleconnection'~\cite{CGT,CGT_Example} in the Northern Hemisphere.
For example, the summer rainfall variation of North America stems from East Asian subtropical monsoon heating by the Asia–North
America teleconnection,~\cite{ANA} as meta-communities covering both Meiyu/Baiu region and parts of North America in~\hyperref[TRMM_GPMAST_CPDS_ES_2000_To_2019_JJA_900_d10_P995_1X1_T2691013G15]{Fig.}~\ref{TRMM_GPMAST_CPDS_ES_2000_To_2019_JJA_900_d10_P995_1X1_T2691013G15}; the positive phase of circumglobal wave train is associated with significantly enhanced precipitation over western Europe and northwestern India,~\cite{CGT,Positive_CGT_Eur} as in~\hyperref[TRMM_GPMAST_CPDS_ES_2000_To_2019_JJA_900_d10_P995_1X1_T18G2]{Fig.}~\ref{TRMM_GPMAST_CPDS_ES_2000_To_2019_JJA_900_d10_P995_1X1_T18G2}.
A detailed case study on this is provided in the next section.

\begin{figure*}[t]
    \centering
    \includegraphics[width=.95\textwidth]{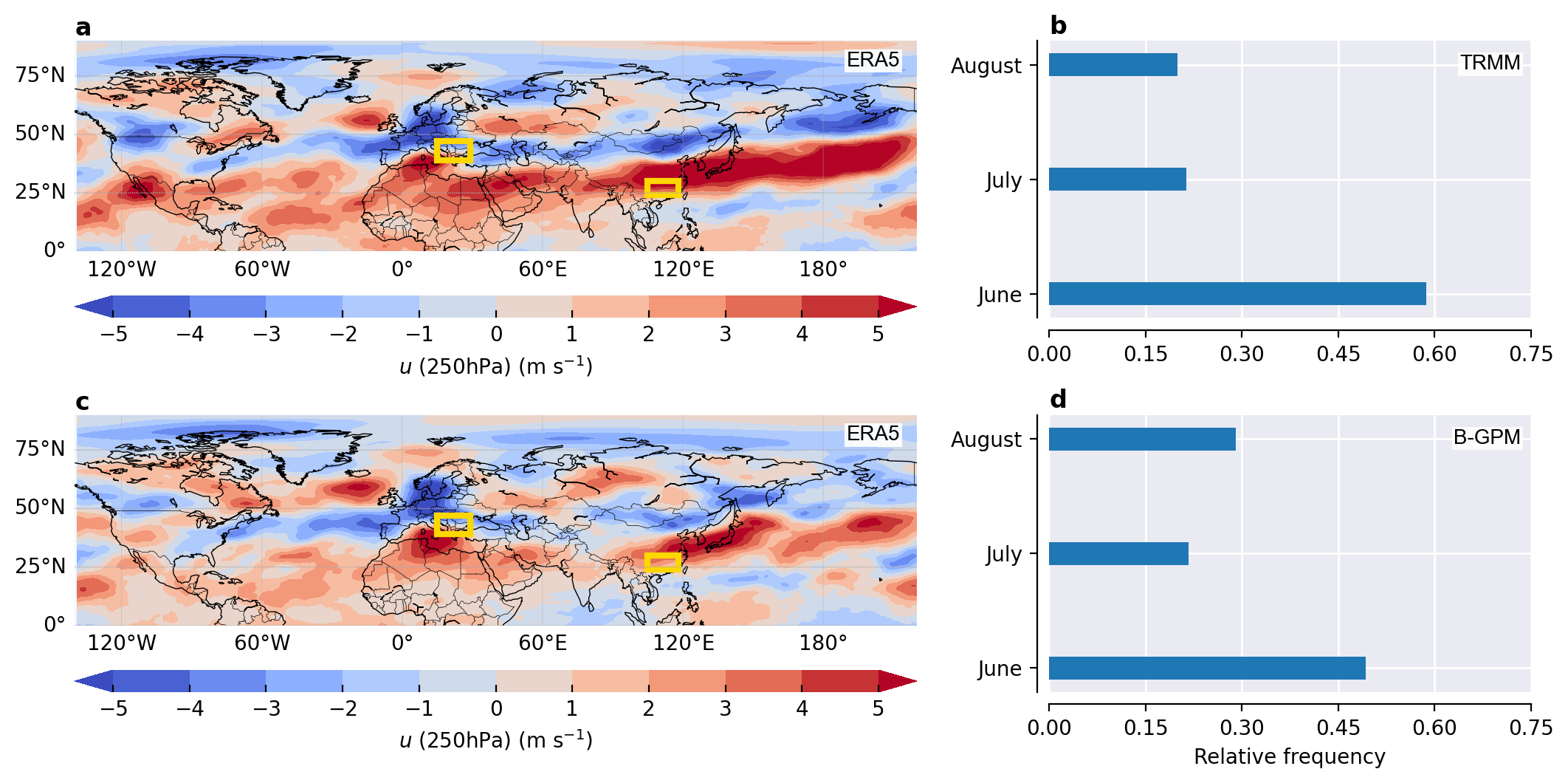}
    \caption{Atmospheric conditions for the teleconnection pattern between South-East Europe and South China, marked with yellow boxes. (a) and (c): Composite anomalies of 250-hPa zonal wind component $u$, with respect to the JJA climatology, based on days when the two regions are highly synchronized (see Sec.~\ref{Atmospheric condition analysis}). For (a) and (c), the days are obtained based on TRMM and B-GPM, respectively. (b) and (d): Frequency of these days over JJA.}
    \label{ES_2000_To_2019_JJA_900_d10_P995_MSSD_ERA5250UCAno_sBALKANS_sSCN1_both_Lag0_CF10}
\end{figure*}

\begin{figure*}[!htbp]
    \centering
    \includegraphics[width=.95\textwidth]{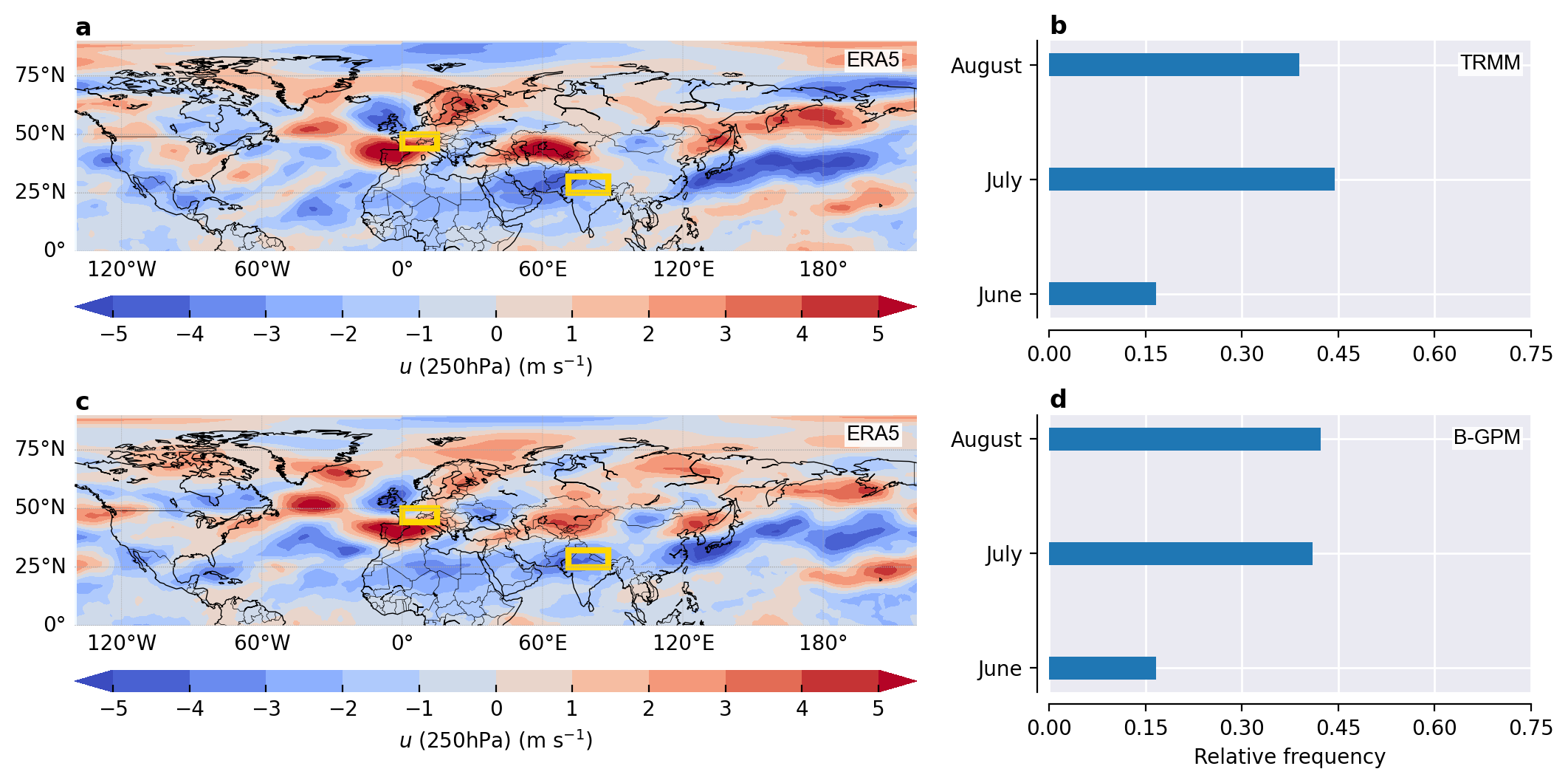}
    \caption{Same as~\hyperref[ES_2000_To_2019_JJA_900_d10_P995_MSSD_ERA5250UCAno_sBALKANS_sSCN1_both_Lag0_CF10]{Fig.}~\ref{ES_2000_To_2019_JJA_900_d10_P995_MSSD_ERA5250UCAno_sBALKANS_sSCN1_both_Lag0_CF10}, but between western and central Europe and northern India.}
    \label{ES_2000_To_2019_JJA_900_d10_P995_MSSD_ERA5250UCAno_sWCEU_sNISM_both_Lag0_CF10}
\end{figure*}

\subsection{Regional significant interdependence of EREs: a case study related to ASM}\label{Regional significant interdependence of EREs: a case study related to ASM}
To further verify the intraseasonal change given by the identified meta-communities, as in~\hyperref[TRMM_GPMAST_CPDS_ES_2000_To_2019_JJA_900_d10_P995_1X1_T2691013G15]{Figs.}~\ref{TRMM_GPMAST_CPDS_ES_2000_To_2019_JJA_900_d10_P995_1X1_T2691013G15} and~\ref{TRMM_GPMAST_CPDS_ES_2000_To_2019_JJA_900_d10_P995_1X1_T18G2}, we here present an ASM-related case study on the upper-level atmospheric circulation.
Two pairs of relationships are chosen, based on significant connections within meta-communities (see~\hyperref[ESLBCSC_TRMM_GPAST_d10_P995_sBALKANS]{Figs.}~\ref{ESLBCSC_TRMM_GPAST_d10_P995_sBALKANS} and~\ref{ESLBCSC_TRMM_GPAST_d10_P995_sNISM} in~\cref{THE PROCESS TO DETERMINE REGIONAL SIGNIFICANT INTERDEPENDENCE FOR THE ASM-RELATED CASE STUDY}).
A particular reason for the first pair in~\hyperref[ESLBCSC_TRMM_GPAST_d10_P995_sBALKANS]{Figs.}~\ref{ESLBCSC_TRMM_GPAST_d10_P995_sBALKANS}\hyperref[ESLBCSC_TRMM_GPAST_d10_P995_sBALKANS]{(e)} and~\ref{ESLBCSC_TRMM_GPAST_d10_P995_sBALKANS}\hyperref[ESLBCSC_TRMM_GPAST_d10_P995_sBALKANS]{(f)} is that, in the early summer of 2018, rainfall extremes occurred in South-East Europe and South-West Japan nearly simultaneously.~\cite{CGT_Example}
Instead of southwestern Japan, we select a large area covering parts of southern China and the Yangtze River Valley, since they are all within the Meiyu/Baiu rainband.
Meanwhile, it has also been suggested that a strong Indian summer monsoon is often accompanied by significant above normal rainfall over part of western and central Europe.~\cite{ES_Teleconnection,Indian_To_Europe}
We select the second pair based on this as indicated in~\hyperref[ESLBCSC_TRMM_GPAST_d10_P995_sNISM]{Figs.}~\ref{ESLBCSC_TRMM_GPAST_d10_P995_sNISM}\hyperref[ESLBCSC_TRMM_GPAST_d10_P995_sNISM]{(e)} and~\ref{ESLBCSC_TRMM_GPAST_d10_P995_sNISM}\hyperref[ESLBCSC_TRMM_GPAST_d10_P995_sNISM]{(f)}.
The composite anomalies, obtained according to the two pairs of relationships, are given below in Figs.~\ref{ES_2000_To_2019_JJA_900_d10_P995_MSSD_ERA5250UCAno_sBALKANS_sSCN1_both_Lag0_CF10} and~\ref{ES_2000_To_2019_JJA_900_d10_P995_MSSD_ERA5250UCAno_sWCEU_sNISM_both_Lag0_CF10}.

In~\hyperref[ES_2000_To_2019_JJA_900_d10_P995_MSSD_ERA5250UCAno_sBALKANS_sSCN1_both_Lag0_CF10]{Figs.}~\ref{ES_2000_To_2019_JJA_900_d10_P995_MSSD_ERA5250UCAno_sBALKANS_sSCN1_both_Lag0_CF10}\hyperref[ES_2000_To_2019_JJA_900_d10_P995_MSSD_ERA5250UCAno_sBALKANS_sSCN1_both_Lag0_CF10]{(b)} and~\ref{ES_2000_To_2019_JJA_900_d10_P995_MSSD_ERA5250UCAno_sBALKANS_sSCN1_both_Lag0_CF10}\hyperref[ES_2000_To_2019_JJA_900_d10_P995_MSSD_ERA5250UCAno_sBALKANS_sSCN1_both_Lag0_CF10]{(d)}, more synchronizations appear in June followed by a decreasing trend, while~\hyperref[ES_2000_To_2019_JJA_900_d10_P995_MSSD_ERA5250UCAno_sWCEU_sNISM_both_Lag0_CF10]{Figs.}~\ref{ES_2000_To_2019_JJA_900_d10_P995_MSSD_ERA5250UCAno_sWCEU_sNISM_both_Lag0_CF10}\hyperref[ES_2000_To_2019_JJA_900_d10_P995_MSSD_ERA5250UCAno_sWCEU_sNISM_both_Lag0_CF10]{(b)} and~\ref{ES_2000_To_2019_JJA_900_d10_P995_MSSD_ERA5250UCAno_sWCEU_sNISM_both_Lag0_CF10}\hyperref[ES_2000_To_2019_JJA_900_d10_P995_MSSD_ERA5250UCAno_sWCEU_sNISM_both_Lag0_CF10]{(d)} present a reversed tendency.
There is a recognizable replacement of an eastward upper-level zonal wind movement, as in~\hyperref[ES_2000_To_2019_JJA_900_d10_P995_MSSD_ERA5250UCAno_sBALKANS_sSCN1_both_Lag0_CF10]{Figs.}~\ref{ES_2000_To_2019_JJA_900_d10_P995_MSSD_ERA5250UCAno_sBALKANS_sSCN1_both_Lag0_CF10}\hyperref[ES_2000_To_2019_JJA_900_d10_P995_MSSD_ERA5250UCAno_sBALKANS_sSCN1_both_Lag0_CF10]{(a)} and~\ref{ES_2000_To_2019_JJA_900_d10_P995_MSSD_ERA5250UCAno_sBALKANS_sSCN1_both_Lag0_CF10}\hyperref[ES_2000_To_2019_JJA_900_d10_P995_MSSD_ERA5250UCAno_sBALKANS_sSCN1_both_Lag0_CF10]{(c)}, with a westward direction, as in~\hyperref[ES_2000_To_2019_JJA_900_d10_P995_MSSD_ERA5250UCAno_sWCEU_sNISM_both_Lag0_CF10]{Figs.}~\ref{ES_2000_To_2019_JJA_900_d10_P995_MSSD_ERA5250UCAno_sWCEU_sNISM_both_Lag0_CF10}\hyperref[ES_2000_To_2019_JJA_900_d10_P995_MSSD_ERA5250UCAno_sWCEU_sNISM_both_Lag0_CF10]{(a)} and~\ref{ES_2000_To_2019_JJA_900_d10_P995_MSSD_ERA5250UCAno_sWCEU_sNISM_both_Lag0_CF10}\hyperref[ES_2000_To_2019_JJA_900_d10_P995_MSSD_ERA5250UCAno_sWCEU_sNISM_both_Lag0_CF10]{(c)} over Japan, for example.
This is consistent with the end of the Meiyu, indicating the disappearance of the high-altitude westerly jet streams and the appearance of the easterlies over the same region.~\cite{EASM_Overview}
Additionally, the upper-level meridional wind movement, as shown in ~\hyperref[ES_2000_To_2019_JJA_900_d10_P995_MSSD_ERA5250VCAno_sBALKANS_sSCN1_both_Lag0_CF10]{Figs.}~\ref{ES_2000_To_2019_JJA_900_d10_P995_MSSD_ERA5250VCAno_sBALKANS_sSCN1_both_Lag0_CF10} and~\ref{ES_2000_To_2019_JJA_900_d10_P995_MSSD_ERA5250VCAno_sWCEU_sNISM_both_Lag0_CF10} in~\cref{THE MERIDIONAL COMPONENT FOR THE ASM-RELATED CASE STUDY}, confirms the function of Rossby waves in creating synchronized EREs for remote mid-latitude regions.~\cite{ES_Teleconnection,SR,CGT}
The identified memberships of two pairs of significant connections to different meta-communities are in agreement with the temporal order and upper-level atmospheric conditions between them.
This case study also suggests that the global synchronization structures in~\hyperref[TRMM_GPMAST_CPDS_ES_2000_To_2019_JJA_900_d10_P995_1X1_T2691013G15]{Figs.}~\ref{TRMM_GPMAST_CPDS_ES_2000_To_2019_JJA_900_d10_P995_1X1_T2691013G15} and~\ref{TRMM_GPMAST_CPDS_ES_2000_To_2019_JJA_900_d10_P995_1X1_T18G2} resemble the distribution of rainfall anomalies (above normal) since only the occurrences of EREs are considered.

\section{Discussion of Robustness}\label{Discussion}
The here proposed clustering workflow enables us to identify regions following similar dynamical properties in an automatic and reliable way.
Parameters, such as $p$, $\gamma$, $n_p$ and $\theta$, are already assigned with appropriate values.
    {This way has confirmed the presence of intraseasonal synchronization structures by using both TRMM and B-GPM data.}
Therefore, we mainly test in the remainder the effects of maximum delay $\tau_{max}$ and the significance level on the identified monsoon-related spatial patterns.

\textit{Robustness over $\tau_{max}$.}
Different choices of maximum delays with $\tau_{max}\in \{3, 30\}$ days are considered, as in~\hyperref[TRMM_GPMAST_LR_d3_P995]{Figs.}~\ref{TRMM_GPMAST_LR_d3_P995} and~\ref{TRMM_GPMAST_LR_d30_P995} in~\cref{THE ROBUSTNESS TEST ON DIFFERENT DELAYS}.
When $\tau_{max}$ decreases (increases) to 3 (30) days, the obtained monsoon-related spatial distribution shrinks (extends) to relatively local (wide) areas (see~\hyperref[TRMM_GPMAST_CPDS_ES_2000_To_2019_JJA_900_d3_P995_1X1_T269G2]{Figs.}~\ref{TRMM_GPMAST_CPDS_ES_2000_To_2019_JJA_900_d3_P995_1X1_T269G2}-\ref{TRMM_GPMAST_CPDS_ES_2000_To_2019_JJA_900_d30_P995_1X1_T26T29} in~\cref{THE ROBUSTNESS TEST ON DIFFERENT DELAYS}).
Indeed, a shorter delay indicates that only those closely synchronized event pairs in time are counted.
These pairs are more likely to be found in neighborhood regions.
The assumption of a short delay might be too strong to capture a long-distance relationship established by the physical atmospheric process.
However, it is also probable that such a relationship appears as the result of a chain of mechanisms with indirect transitivity.
This can be revealed from the comparisons between~\hyperref[TRMM_GPMAST_CPDS_ES_2000_To_2019_JJA_900_d3_P995_1X1_T269G2]{Figs.}~\ref{TRMM_GPMAST_CPDS_ES_2000_To_2019_JJA_900_d3_P995_1X1_T269G2} and~\ref{TRMM_GPMAST_CPDS_ES_2000_To_2019_JJA_900_d30_P995_1X1_T13451011G135710}, or~\hyperref[TRMM_GPMAST_CPDS_ES_2000_To_2019_JJA_900_d3_P995_1X1_T1G1]{Figs.}~\ref{TRMM_GPMAST_CPDS_ES_2000_To_2019_JJA_900_d3_P995_1X1_T1G1} and~\ref{TRMM_GPMAST_CPDS_ES_2000_To_2019_JJA_900_d30_P995_1X1_T26T29}, since the spatial pattern under the delay of 3 days is preserved under that of 30 days.
That is, the identified intraseasonal synchronization structures are stable inside the interval $\tau_{max} \in [3, 30]$.

\textit{Robustness over significance level.}
The significance level determines the topological structure of the reconstructed functional climate network.
A particularly low threshold may eliminate some important synchronization relationships,~\cite{ES_Teleconnection} while a higher value amplifies the multiple comparison bias.
We therefore consider the conservative choices of $P<0.006$ and $P<0.004$ (see~\hyperref[TRMM_GPMAST_LR_d10_P994]{Figs.}~\ref{TRMM_GPMAST_LR_d10_P994} and~\ref{TRMM_GPMAST_LR_d10_P996} in~\cref{THE ROBUSTNESS TEST ON DIFFERENT SIGNIFICANCE LEVELS}).
The monsoon-related spatial distributions, shown in~\hyperref[TRMM_GPMAST_CPDS_ES_2000_To_2019_JJA_900_d10_P994_1X1_T2691013G15]{Figs.}~\ref{TRMM_GPMAST_CPDS_ES_2000_To_2019_JJA_900_d10_P994_1X1_T2691013G15}-\ref{TRMM_GPMAST_CPDS_ES_2000_To_2019_JJA_900_d10_P996_1X1_T110G2} in~\cref{THE ROBUSTNESS TEST ON DIFFERENT SIGNIFICANCE LEVELS}, remain consistent with~\hyperref[TRMM_GPMAST_CPDS_ES_2000_To_2019_JJA_900_d10_P995_1X1_T2691013G15]{Figs.}~\ref{TRMM_GPMAST_CPDS_ES_2000_To_2019_JJA_900_d10_P995_1X1_T2691013G15} and~\ref{TRMM_GPMAST_CPDS_ES_2000_To_2019_JJA_900_d10_P995_1X1_T18G2}.
By means of the comparison between~\hyperref[TRMM_GPMAST_LR_d10_P995]{Figs.}~\ref{TRMM_GPMAST_LR_d10_P995} and~\ref{TRMM_GPMAST_LR_d10_P994}, or~\hyperref[TRMM_GPMAST_LR_d10_P996]{Figs.}~\ref{TRMM_GPMAST_LR_d10_P996} and~\ref{TRMM_GPMAST_LR_d10_P994}, lowering the significance threshold yields a slight increase in modularity.
Such a signal implies the community structure as an intrinsic attribute of the underlying self-organized climate system.

    {In spite of the robustness of the identified two synchronization structures, it is still worth noting that several regions remain dissimilar to a certain degree.
        For example, the meta-community in~\hyperref[TRMM_GPMAST_CPDS_ES_2000_To_2019_JJA_900_d10_P995_1X1_T2691013G15]{Fig.}~\ref{TRMM_GPMAST_CPDS_ES_2000_To_2019_JJA_900_d10_P995_1X1_T2691013G15}\hyperref[TRMM_GPMAST_CPDS_ES_2000_To_2019_JJA_900_d10_P995_1X1_T2691013G15]{(c)} covers a larger area of the North Atlantic than that in~\hyperref[TRMM_GPMAST_CPDS_ES_2000_To_2019_JJA_900_d10_P995_1X1_T2691013G15]{Fig.}~\ref{TRMM_GPMAST_CPDS_ES_2000_To_2019_JJA_900_d10_P995_1X1_T2691013G15}\hyperref[TRMM_GPMAST_CPDS_ES_2000_To_2019_JJA_900_d10_P995_1X1_T2691013G15]{(a)}.
        We attribute this to the uncertainties from rainfall data, since fluctuations caused by the network-based clustering approach are corrected in our proposed workflow.
        Ideally, when two precipitation datasets are exactly the same, the reconstructed network structures are identical and the resulting spatial distributions can be perfectly matched.
        However, satellite observations contain non-negligible random errors and biases during the generation of datasets, due to factors such as inadequate sampling and deficiencies in algorithms.~\cite{Uncertainties}
        These uncertainties are further retained, leading to discrepancies in the identified synchronization structures.
        However, the discrepancy information can also be important to measure the extent to which different datasets reproduce the same climate phenomenon.}

\section{Conclusion}\label{Conclusion}
In summary, our work starts from the identification of regions of similar climatological behavior and focuses on revealing the global view of the interdependence patterns for EREs in the context of the global monsoon.
For this, we provide a network-based clustering workflow that combines consensus clustering and mutual correspondences.
By means of this clustering workflow, we identify two main monsoon-related synchronization structures.
They have independent temporal and spatial characteristics.
The first structure, primarily from early-June to mid-July, shows early summer features over different monsoon systems, including the synchronized initiation of the Indian rainy season and Meiyu/Baiu over ASM, the oceanic regime over West Africa, and early-rainy season over the Caribbean Sea towards the central United States.
The second structure, from mid-July to late-August, manifests the states after the `monsoon jump', including the establishment of the `north' teleconnection inside the ASM between Indian summer monsoon and the rainfall in North China, and the continental regime over West Africa.
Meanwhile, the mid-summer drought is dominating Central America and southern Mexico.
Also, we show that the significant connections inside the identified meta-communities capture the synchronized behavior via the `Silk Road teleconnection'~\cite{SR} and the `circumglobal teleconnection'~\cite{CGT,CGT_Example} for the mid-latitude belt.

The successful identification of intraseasonal spatiotemporal variability essentially stems from the combination of ES and complex-network theory to represent the underlying climate system as network.
This clustering workflow can be also applied to other seasonal analyses like the boreal winter, which brings marked monsoonal characteristics to the Southern Hemisphere.
Also, the extension to a higher spatial resolution will be investigated in the near future.

\begin{acknowledgments}
    We would like to thank Shraddha Gupta, Niklas Boers, and Frederik Wolf for valuable discussions as well as Yang Liu and Fabian Brandt-Tumescheit for technical support.
    This work was financially supported by the China Scholarship Council (CSC)
    scholarship.
\end{acknowledgments}

\appendix
\clearpage
\onecolumngrid
\renewcommand{\appendixname}{APPENDIX}
\section{THE MOTIVATION FOR THE NETWORK-BASED WORKFLOW}\label[Appendix]{THE MOTIVATION FOR THE NETWORK-BASED WORKFLOW}

\begin{figure*}[!htbp]
    \centering
    \includegraphics[width=.95\textwidth]{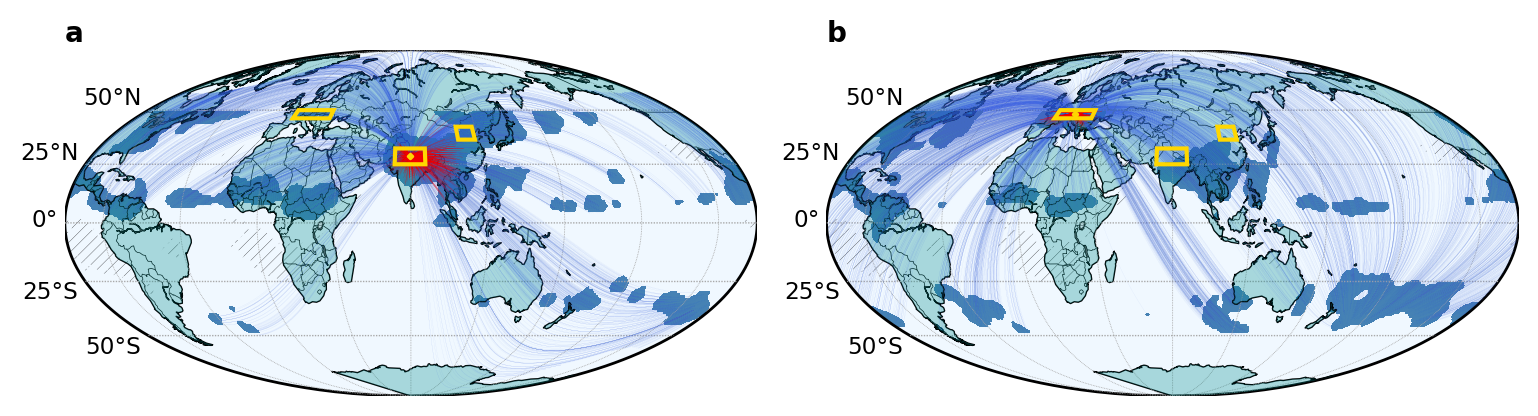}
    \caption{Illustration of the limitation of the significant test for link bundles under TRMM. Three yellow boxes are positioned over Europe, northern India, and northern China, respectively. (a) When the box in northern India is selected as the central region, associated significant link bundles connect to {the entire boxes in both Europe and northern China.}
        (b) When the central region is changed to the box in Europe, only a few link bundles connect to the boxes in northern India and northern China. This indicates that the significant link bundles are limited by the link distribution of a given region, and therefore can only reveal a partial view of regions that follow similar climatological behavior.}
    \label[Fig.]{ESLB_1998_To_2019_JJA_900_d10_P995_LBP999_X4_Exp_Tele}
\end{figure*}

\begin{figure*}[!htbp]
    \centering
    \includegraphics[width=.65\textwidth]{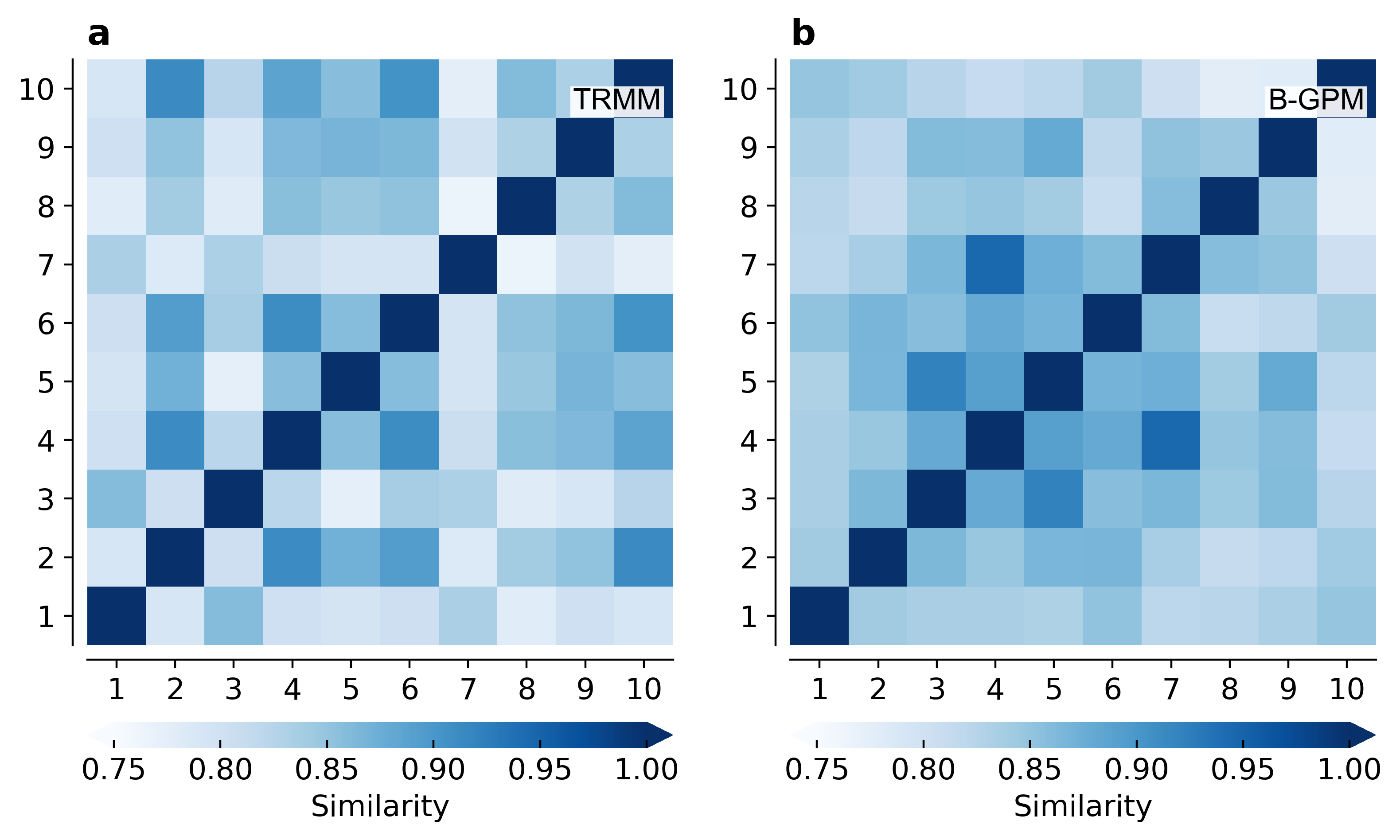}
    \caption{Illustration of fluctuation of clustering approaches, such as PLM, under (a) TRMM and (b) B-GPM, respectively. Taking TRMM as an example, the first 10 partitions of the highest modularity are selected from 1000 randomly generated partitions. Any pair between them should show a similarity of 1, based on the normalized mutual information score, if the clustering process were deterministic. However, (a) and (b) show the average deviations of 0.15 and 0.14, respectively, indicating a certain degree of differences between solutions.
    }
    \label{CD_Random_Noise}
\end{figure*}

\clearpage
\section{COMMUNITY STRUCTURE UNDER THE DELAY OF 10 DAYS}\label[Appendix]{COMMUNITY STRUCTURE UNDER THE DELAY OF 10 DAYS}

\begin{figure*}[!htbp]
    \centering
    \includegraphics[width=.85\textwidth]{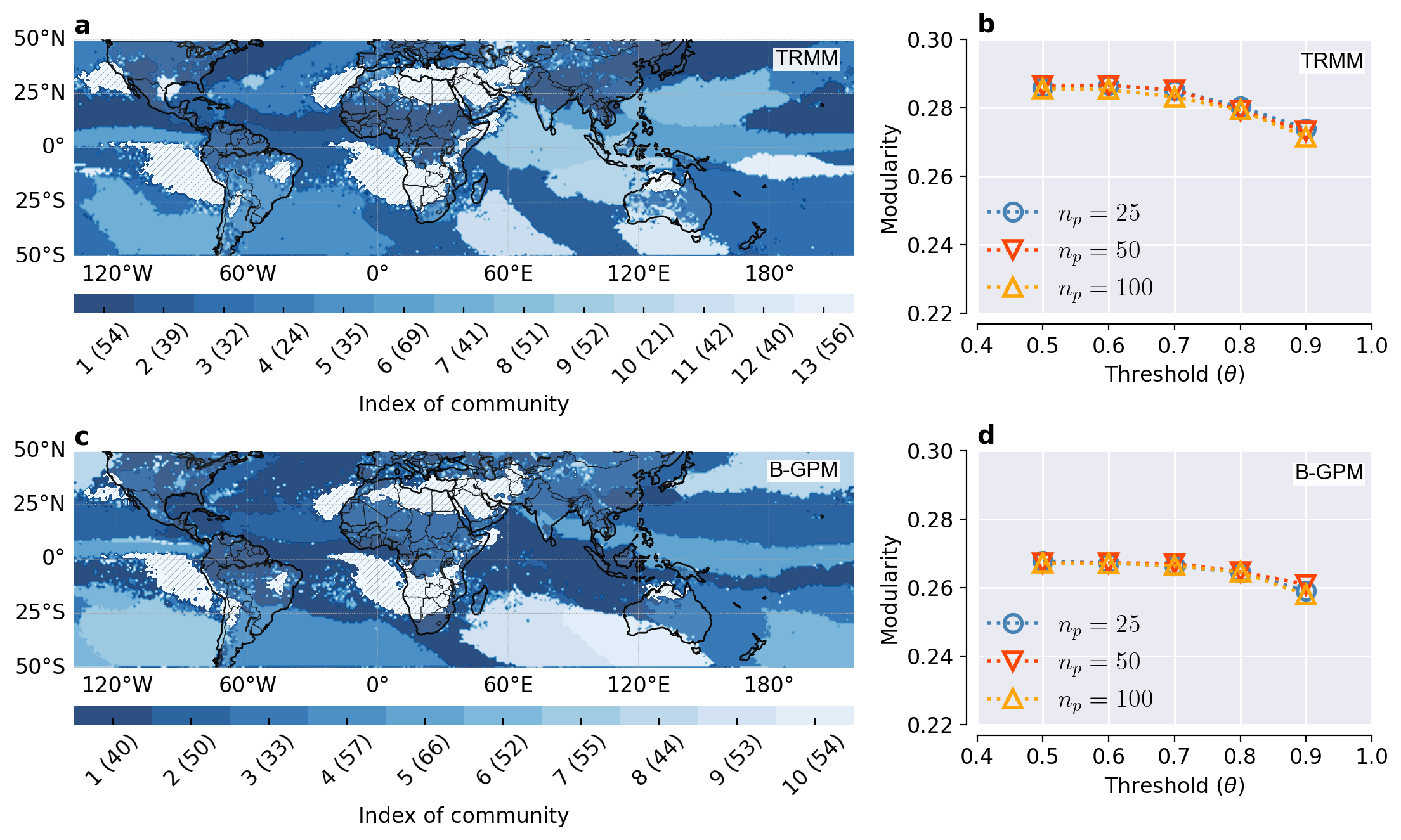}
    \caption{Community structures for TRMM and B-GPM, respectively.
        (b) and (d): Modularity as a function of $n_p$ and $\tau$ for the selection of community structure for TRMM and B-GPM.
        (a) and (c) share a similarity of 0.58, based on normalized mutual information score.
        They are chosen according to (b) with $n_p=25$ and $\theta=0.6$ and (d) with $n_p=25$ and $\theta=0.5$, respectively, to indicate the maximum modularity after consensus clustering.
        Both are sorted by the size of each community, namely, the number of grid points inside of a community.
        Each index is attached with the average number of EREs in the corresponding community.
        Communities smaller than 0.01*$|V|$ are excluded from the visualization. Grid points with less than three events are also excluded when extracting EREs and visualized as hatched areas.
    }
    \label{TRMM_GPMAST_LR_d10_P995}
\end{figure*}

\clearpage
\section{THE PROCESS TO DETERMINE REGIONAL SIGNIFICANT INTERDEPENDENCE FOR THE ASM-RELATED CASE STUDY}\label[Appendix]{THE PROCESS TO DETERMINE REGIONAL SIGNIFICANT INTERDEPENDENCE FOR THE ASM-RELATED CASE STUDY}

\begin{figure*}[!htbp]
    \centering
    \includegraphics[width=.95\textwidth]{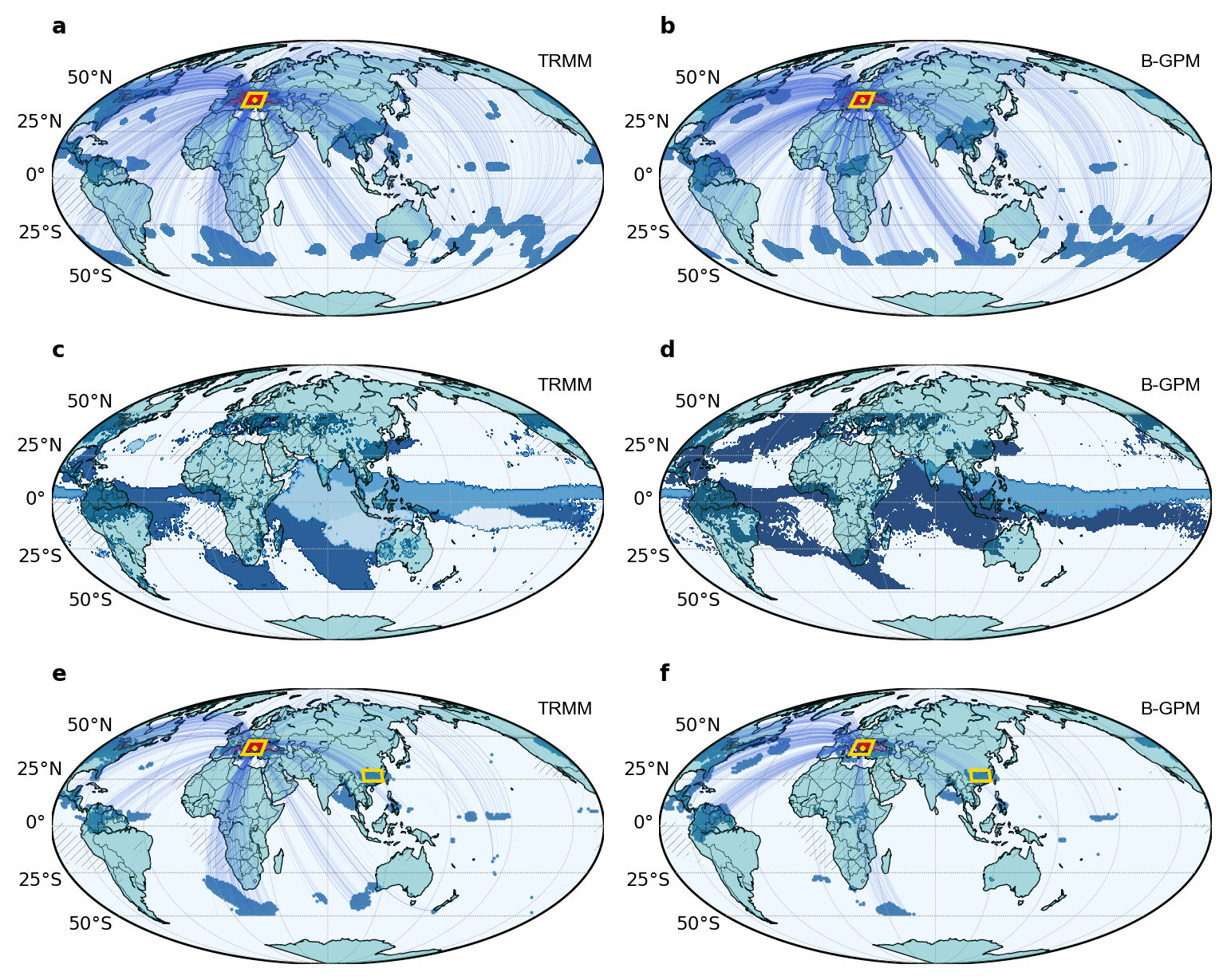}
    \caption{Teleconnection pattern given for South-East Europe, for TRMM ((a), (c), and (e)) and B-GPM ((b), (d), and (f)). (a) and (b): Significant link bundles as given in Ref.~\onlinecite{ES_Teleconnection} but for South-East Europe (see Sec.~\ref{The network-based clustering workflow} `Significance test for link bundles'). (c) and (d): Meta-communities extracted from~\hyperref[TRMM_GPMAST_LR_d10_P995]{Fig.}~\ref{TRMM_GPMAST_LR_d10_P995} and same as~\hyperref[TRMM_GPMAST_CPDS_ES_2000_To_2019_JJA_900_d10_P995_1X1_T2691013G15]{Fig.}~\ref{TRMM_GPMAST_CPDS_ES_2000_To_2019_JJA_900_d10_P995_1X1_T2691013G15} (see Sec.~\ref{The network-based clustering workflow} `Consensus clustering' and `Mutual correspondences'). (e) and (f): Regional significant interdependence based on the intersections of (a) and (c), and (b) and (d), respectively. Yellow boxes positioned in (e) and (f) represent the presence of significant connections between South-East Europe (39$^{\circ}$N to 47$^{\circ}$N, 15$^{\circ}$E to 29$^{\circ}$E) and South China (24$^{\circ}$N to 30$^{\circ}$N, 105$^{\circ}$E to 118$^{\circ}$E), to be considered in Sec.~\ref{Regional significant interdependence of EREs: a case study related to ASM}.}
    \label{ESLBCSC_TRMM_GPAST_d10_P995_sBALKANS}
\end{figure*}

\clearpage
\begin{figure*}[!htbp]
    \centering
    \includegraphics[width=.95\textwidth]{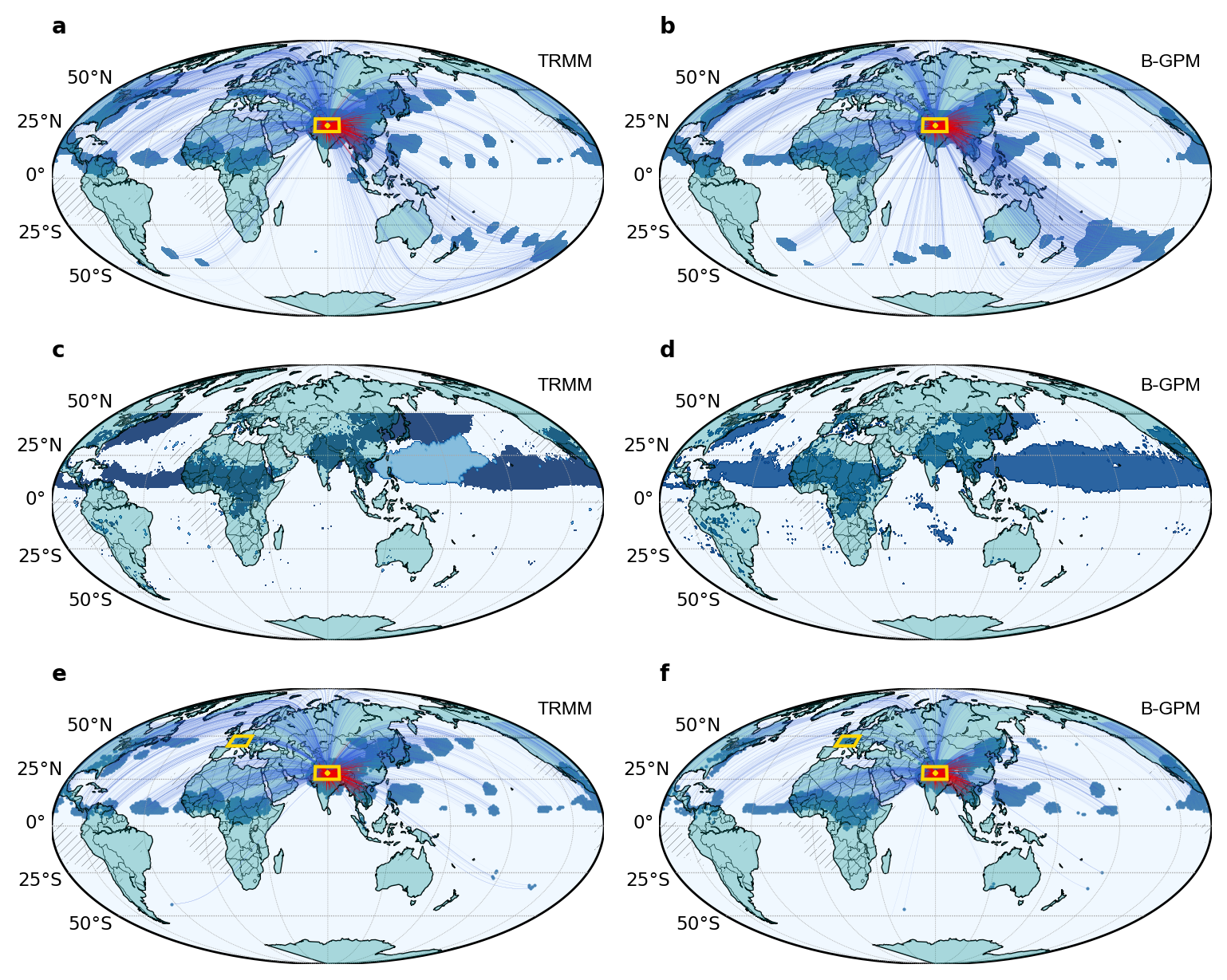}
    \caption{Teleconnection pattern given for Northern India, for TRMM ((a), (c), and (e)) and B-GPM ((b), (d), and (f)). (a) and (b): Significant link bundles as given in Ref.~\onlinecite{ES_Teleconnection} (see Sec.~\ref{The network-based clustering workflow} `Significance test for link bundles'). (c) and (d): Meta-communities extracted from Fig.~\ref{TRMM_GPMAST_LR_d10_P995} and same as~\hyperref[TRMM_GPMAST_CPDS_ES_2000_To_2019_JJA_900_d10_P995_1X1_T18G2]{Fig.}~\ref{TRMM_GPMAST_CPDS_ES_2000_To_2019_JJA_900_d10_P995_1X1_T18G2} (see Sec.~\ref{The network-based clustering workflow} `Consensus clustering' and `Mutual correspondences'). (e) and (f): Regional significant interdependence based on the intersections of (a) and (c), and (b) and (d), respectively. Yellow boxes positioned in (e) and (f) represent the presence of significant connections between Northern India (25$^{\circ}$N to 32$^{\circ}$N, 71$^{\circ}$E to 88$^{\circ}$E) and part of western and central Europe (44$^{\circ}$N to 50$^{\circ}$N, 0$^{\circ}$E to 15$^{\circ}$E), to be considered in Sec.~\ref{Regional significant interdependence of EREs: a case study related to ASM}.}
    \label{ESLBCSC_TRMM_GPAST_d10_P995_sNISM}
\end{figure*}

\clearpage
\section{THE MERIDIONAL COMPONENT FOR THE ASM-RELATED CASE STUDY}\label[Appendix]{THE MERIDIONAL COMPONENT FOR THE ASM-RELATED CASE STUDY}
\begin{figure*}[!htbp]
    \centering
    \includegraphics[width=.95\textwidth]{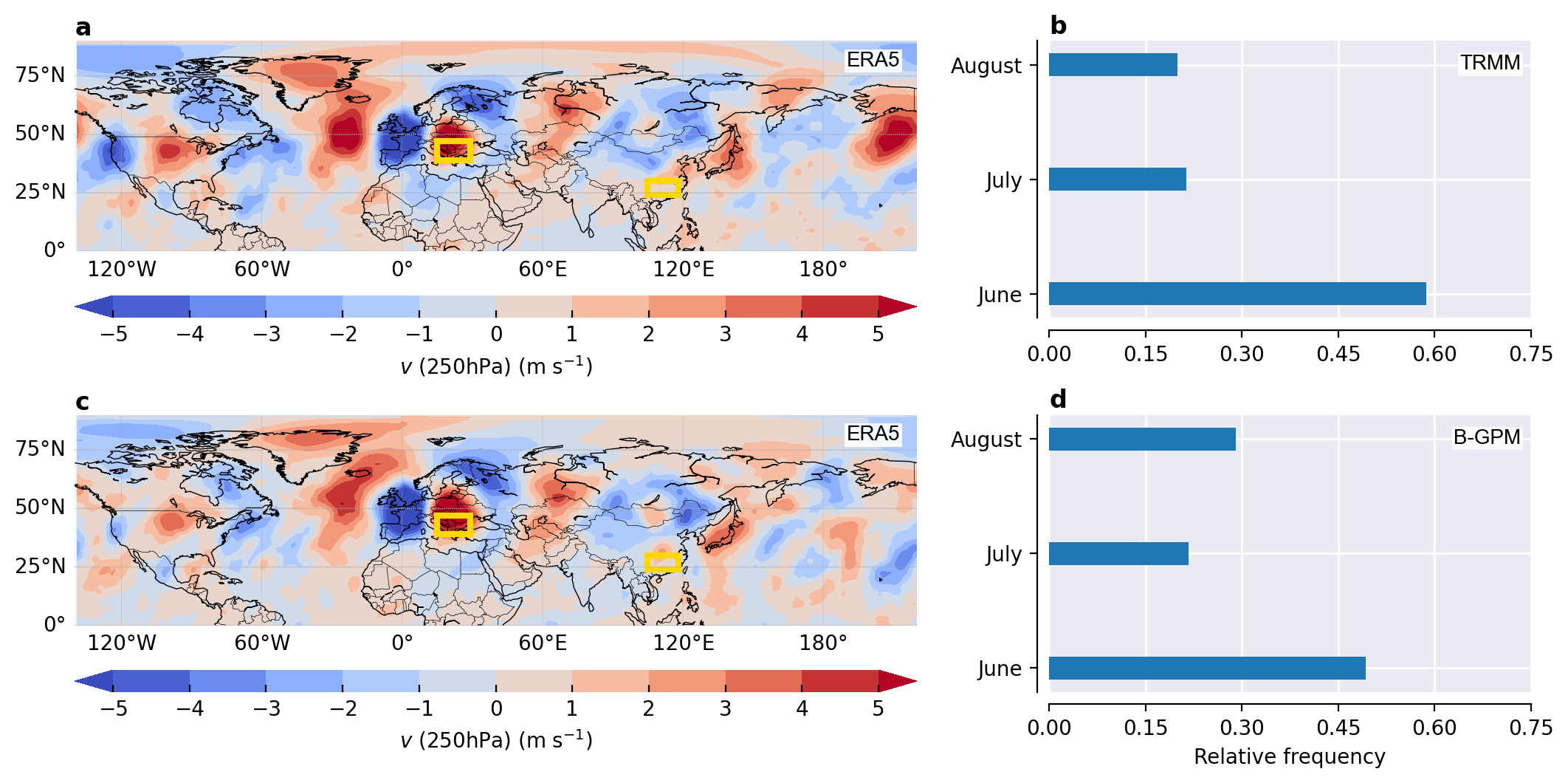}
    \caption{Atmospheric conditions for the teleconnection pattern between South-East Europe and South China, marked with yellow boxes. (a) and (c): Composite anomalies of 250-hPa meridional wind component $v$, with respect to the JJA climatology, based on days when the two regions are highly synchronized (see Sec.~\ref{Atmospheric condition analysis}). For (a) and (c), the days are obtained based on TRMM and B-GPM, respectively. (b) and (d): Frequency of these days over JJA.}
    \label{ES_2000_To_2019_JJA_900_d10_P995_MSSD_ERA5250VCAno_sBALKANS_sSCN1_both_Lag0_CF10}
\end{figure*}

\begin{figure*}[!htbp]
    \centering
    \includegraphics[width=.95\textwidth]{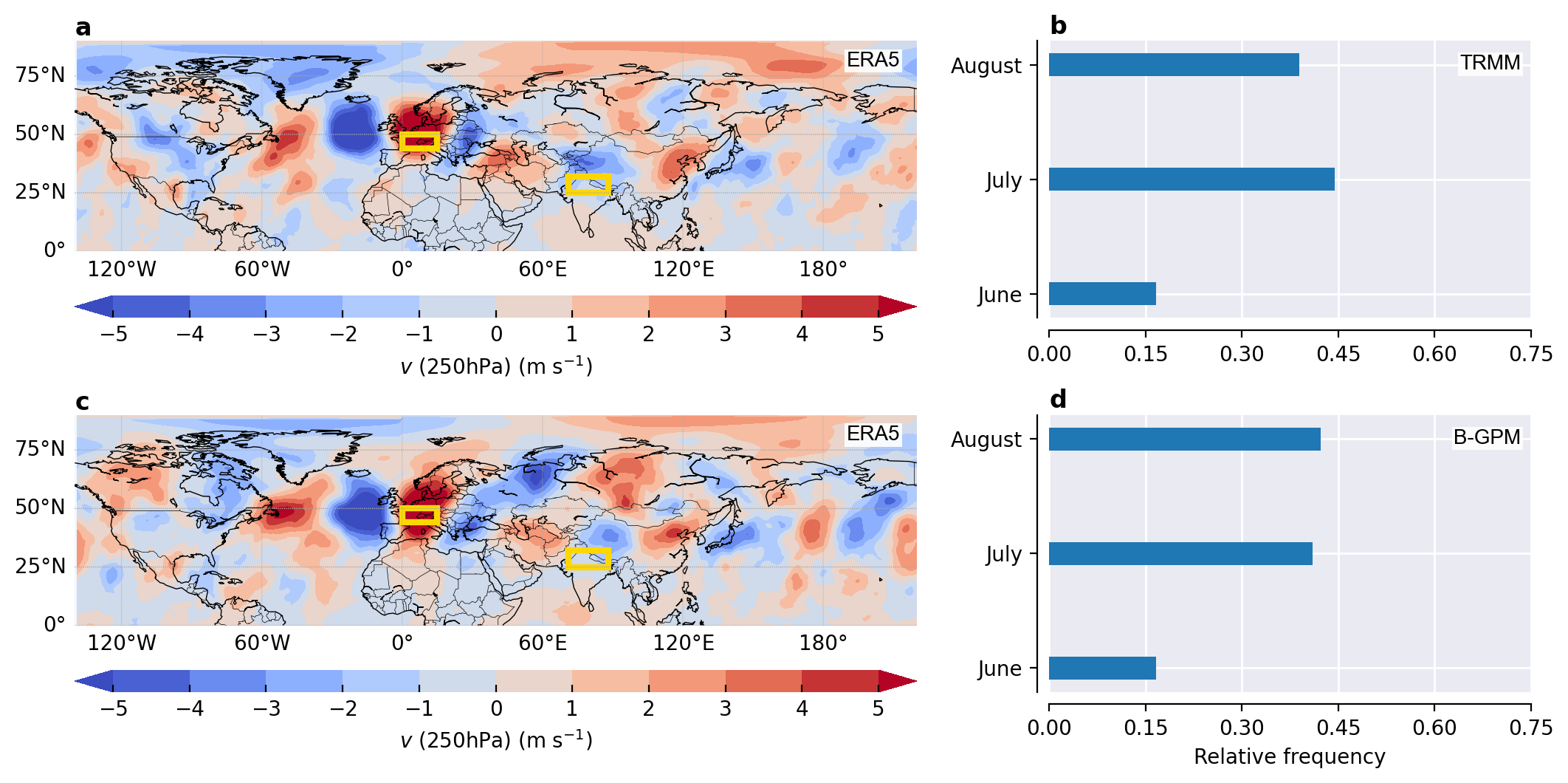}
    \caption{Same as~\hyperref[ES_2000_To_2019_JJA_900_d10_P995_MSSD_ERA5250VCAno_sBALKANS_sSCN1_both_Lag0_CF10]{Fig.}~\ref{ES_2000_To_2019_JJA_900_d10_P995_MSSD_ERA5250VCAno_sBALKANS_sSCN1_both_Lag0_CF10}, but between western and central Europe and northern India.}
    \label{ES_2000_To_2019_JJA_900_d10_P995_MSSD_ERA5250VCAno_sWCEU_sNISM_both_Lag0_CF10}
\end{figure*}

\clearpage
\section{THE ROBUSTNESS TEST ON DIFFERENT DELAYS}\label[Appendix]{THE ROBUSTNESS TEST ON DIFFERENT DELAYS}
\begin{figure*}[!htbp]
    \centering
    \includegraphics[width=.85\textwidth]{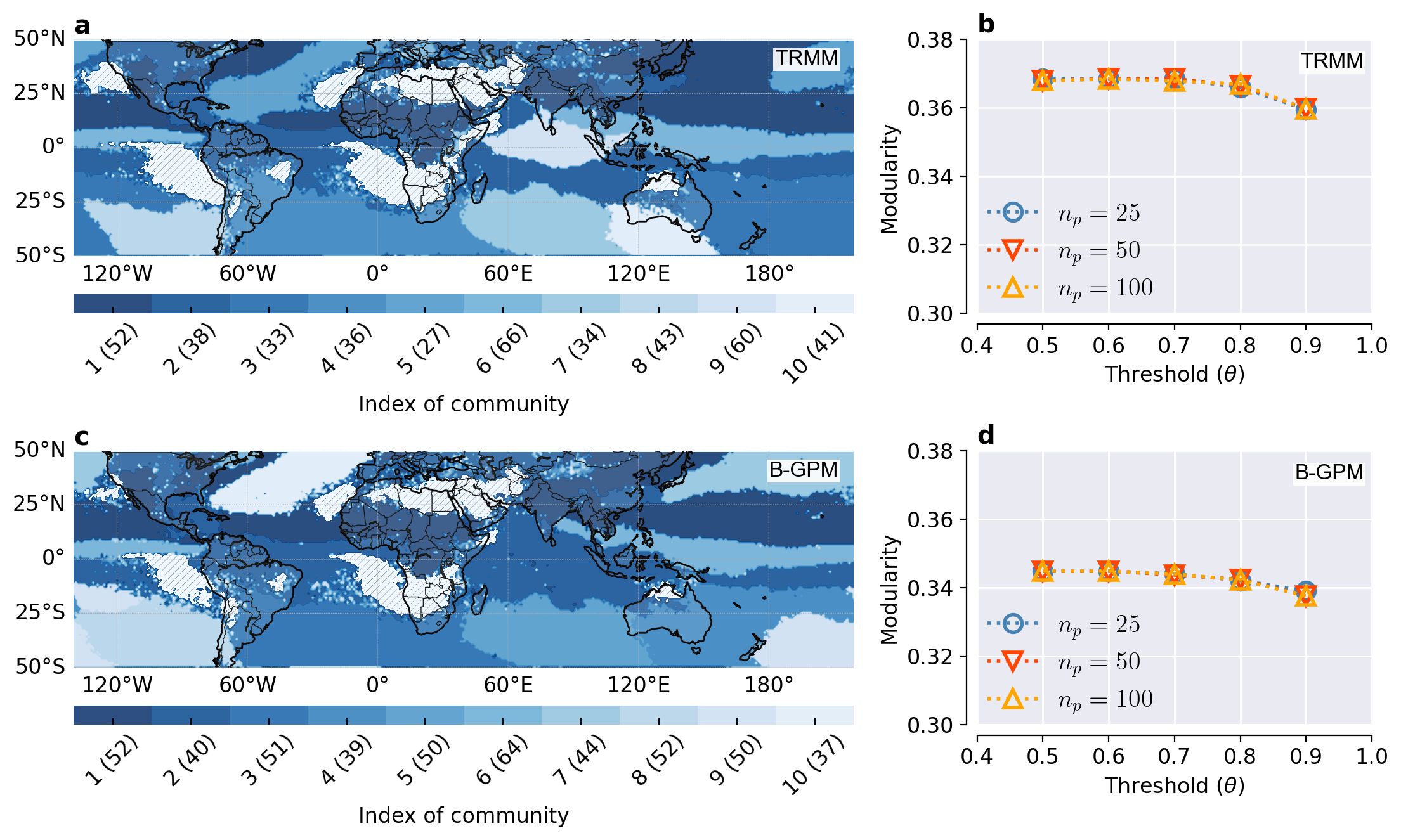}
    \caption{
        Same as~\hyperref[TRMM_GPMAST_LR_d10_P995]{Fig.}~\ref{TRMM_GPMAST_LR_d10_P995}, but for a different $\tau_{max}=3$ days.
        In (a) and (c), community structures for TRMM and B-GPM are chosen according to (b) with $n_p=25$ and $\theta=0.5$ and (d) with $n_p=50$ and $\theta=0.6$, respectively, to indicate the highest modularity value.
        They share a similarity of 0.66, based on the normalized mutual information score.
    }
    \label{TRMM_GPMAST_LR_d3_P995}
\end{figure*}

\begin{figure*}[!htbp]
    \centering
    \includegraphics[width=.85\textwidth]{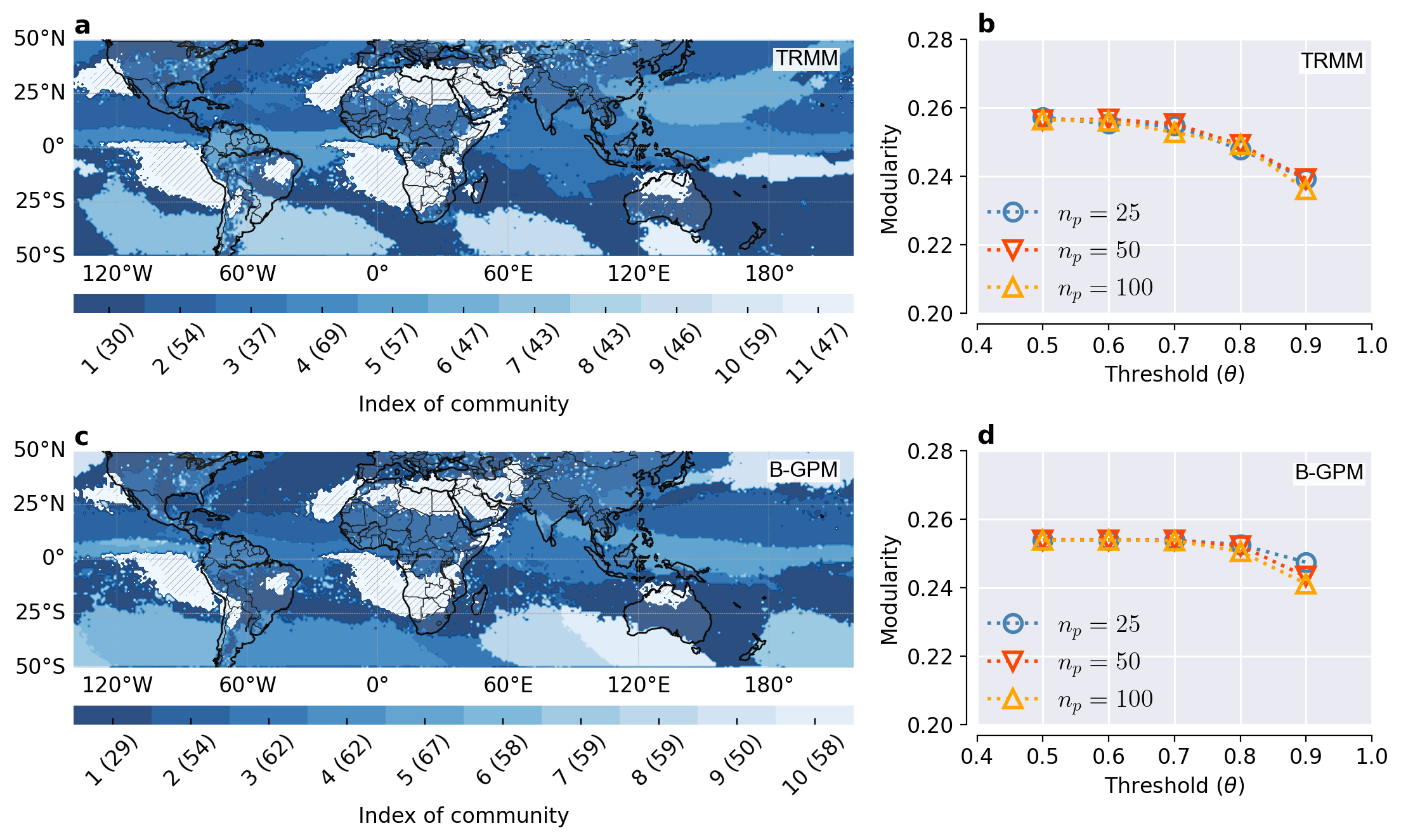}
    \caption{
        Same as~\hyperref[TRMM_GPMAST_LR_d10_P995]{Fig.}~\ref{TRMM_GPMAST_LR_d10_P995}, but for a different $\tau_{max}=30$ days.
        Community structures for (a) TRMM and (c) B-GPM, are chosen according to (b) with $n_p=25$ and $\theta=0.5$ and (d) with $n_p=50$ and $\theta=0.6$, respectively, sharing a similarity of 0.60.
    }
    \label{TRMM_GPMAST_LR_d30_P995}
\end{figure*}

\begin{figure*}[!htbp]
    \centering
    \includegraphics[width=.95\textwidth]{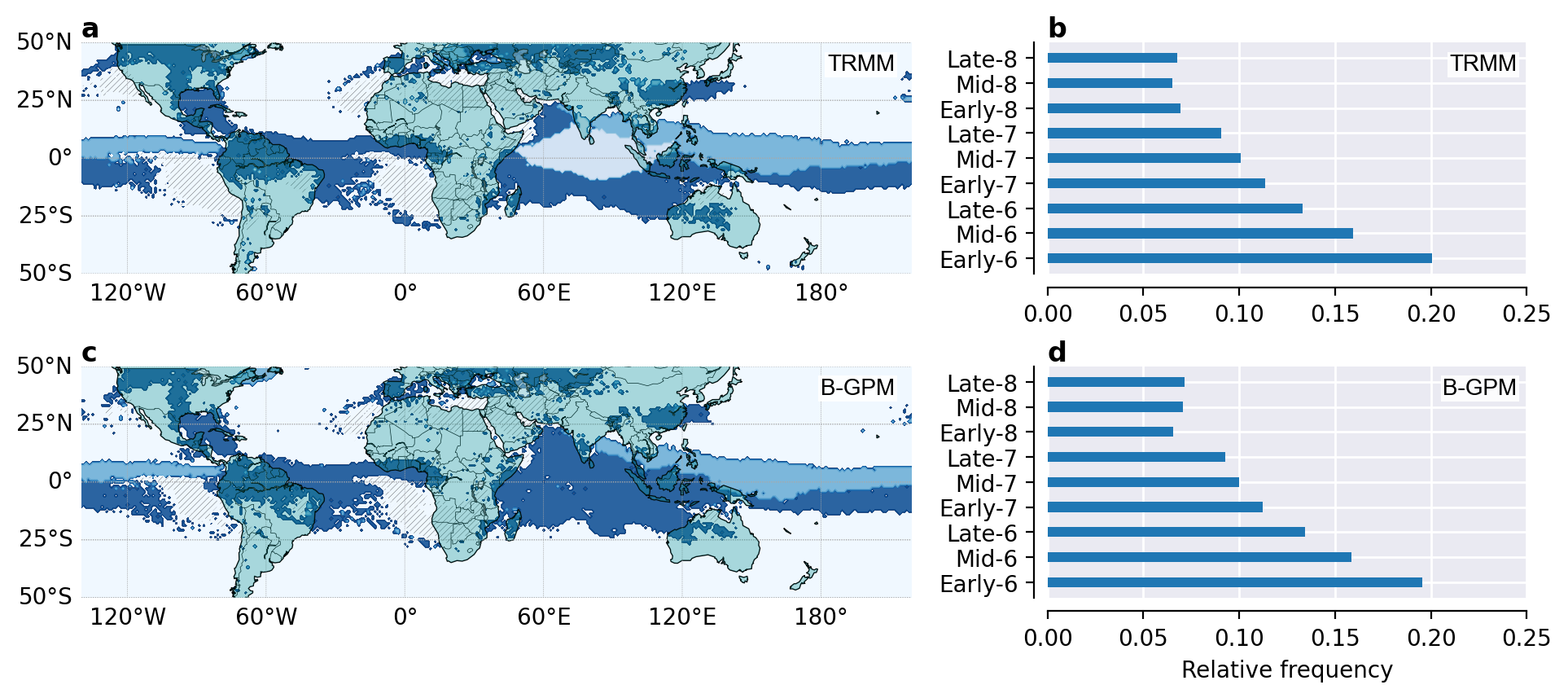}
    \caption{Same as~\hyperref[TRMM_GPMAST_CPDS_ES_2000_To_2019_JJA_900_d10_P995_1X1_T2691013G15]{Fig.}~\ref{TRMM_GPMAST_CPDS_ES_2000_To_2019_JJA_900_d10_P995_1X1_T2691013G15}, but for a different $\tau_{max}=3$ days. Communities of mutual correspondence with a similarity of $RCR($TRMM=\{2, 6, 9\}, B-GPM=\{2, 6\}$)=0.72$. (a) Communities 2, 6, and 9 from~\hyperref[TRMM_GPMAST_LR_d3_P995]{Fig.}~\ref{TRMM_GPMAST_LR_d3_P995}\hyperref[TRMM_GPMAST_LR_d3_P995]{(a)}. (c) Communities 2 and 6 from~\hyperref[TRMM_GPMAST_LR_d3_P995]{Fig.}~\ref{TRMM_GPMAST_LR_d3_P995}\hyperref[TRMM_GPMAST_LR_d3_P995]{(b)}. (b) and (d): Frequency of synchronized days within meta-communities in (a) and (c), respectively (see Sec.~\ref{The network-based clustering workflow} `Identification of synchronized days for meta-communities'). Grid points with less than three events are excluded when extracting EREs and visualized as hatched areas here.
    }
    \label{TRMM_GPMAST_CPDS_ES_2000_To_2019_JJA_900_d3_P995_1X1_T269G2}
\end{figure*}

\begin{figure*}[!htbp]
    \centering
    \includegraphics[width=.95\textwidth]{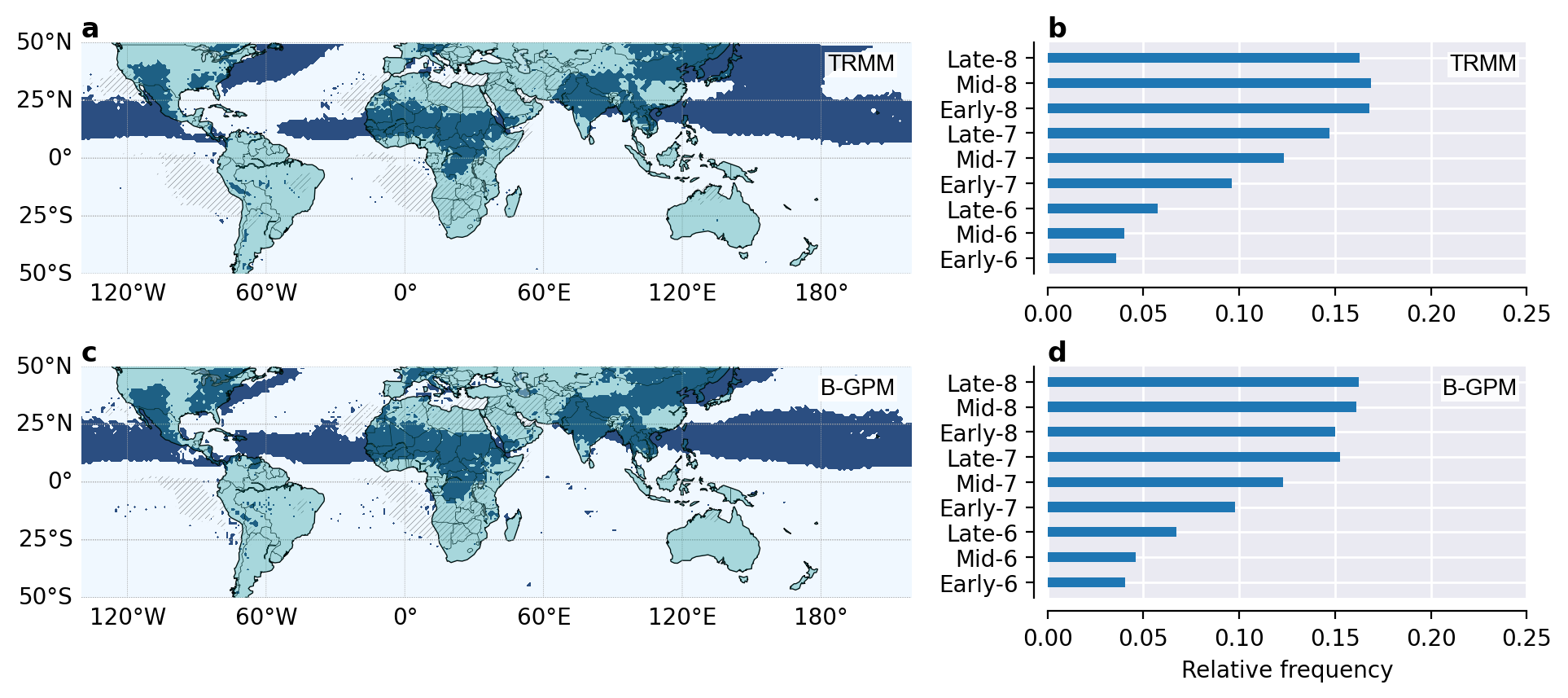}
    \caption{Same as~\hyperref[TRMM_GPMAST_CPDS_ES_2000_To_2019_JJA_900_d10_P995_1X1_T18G2]{Fig.}~\ref{TRMM_GPMAST_CPDS_ES_2000_To_2019_JJA_900_d10_P995_1X1_T18G2}, but for a different $\tau_{max}=3$ days. Communities of mutual correspondence with a similarity of $RCR($TRMM=\{1\}, B-GPM=\{1\}$)=0.65$ from~\hyperref[TRMM_GPMAST_LR_d3_P995]{Fig.}~\ref{TRMM_GPMAST_LR_d3_P995}.
    }
    \label{TRMM_GPMAST_CPDS_ES_2000_To_2019_JJA_900_d3_P995_1X1_T1G1}
\end{figure*}

\begin{figure*}[!htbp]
    \centering
    \includegraphics[width=.95\textwidth]{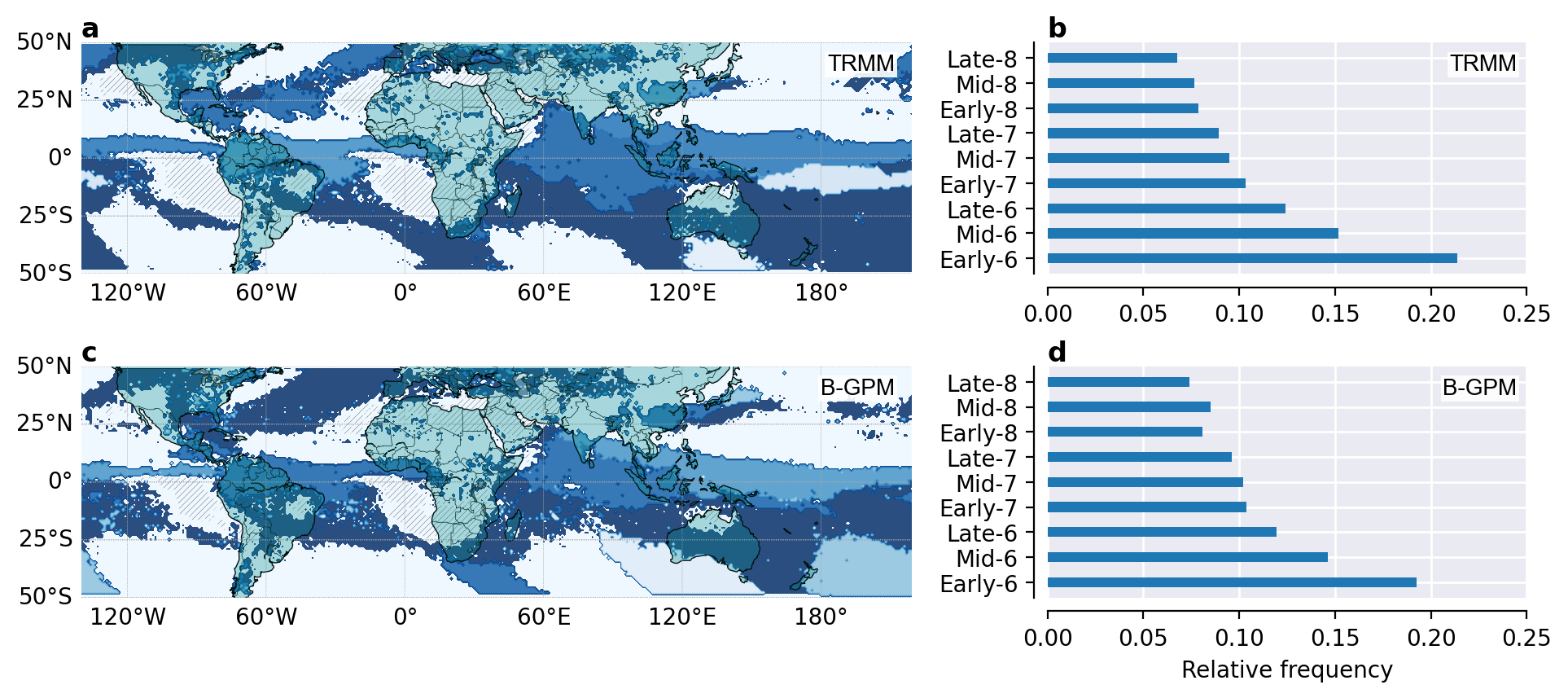}
    \caption{Same as~\hyperref[TRMM_GPMAST_CPDS_ES_2000_To_2019_JJA_900_d10_P995_1X1_T2691013G15]{Fig.}~\ref{TRMM_GPMAST_CPDS_ES_2000_To_2019_JJA_900_d10_P995_1X1_T2691013G15}, but for a different $\tau_{max}=30$ days. Communities of mutual correspondence with a similarity of $RCR($TRMM=\{1, 3, 4, 5, 10, 11\}, B-GPM=\{1, 3, 5, 7, 10\}$)=0.72$. (a) Communities 1, 3, 4, 5, 10 and 11 from~\hyperref[TRMM_GPMAST_LR_d30_P995]{Fig.}~\ref{TRMM_GPMAST_LR_d30_P995}\hyperref[TRMM_GPMAST_LR_d30_P995]{(a)}. (c) Communities 1, 3, 5, 7 and 10 from~\hyperref[TRMM_GPMAST_LR_d30_P995]{Fig.}~\ref{TRMM_GPMAST_LR_d30_P995}\hyperref[TRMM_GPMAST_LR_d30_P995]{(b)}. (b) and (d): Frequency of synchronized days within meta-communities in (a) and (c), respectively (see Sec.~\ref{The network-based clustering workflow} `Identification of synchronized days for meta-communities'). Grid points with less than three events are excluded when extracting EREs and visualized as hatched areas here.
    }
    \label{TRMM_GPMAST_CPDS_ES_2000_To_2019_JJA_900_d30_P995_1X1_T13451011G135710}
\end{figure*}

\begin{figure*}[!htbp]
    \centering
    \includegraphics[width=.95\textwidth]{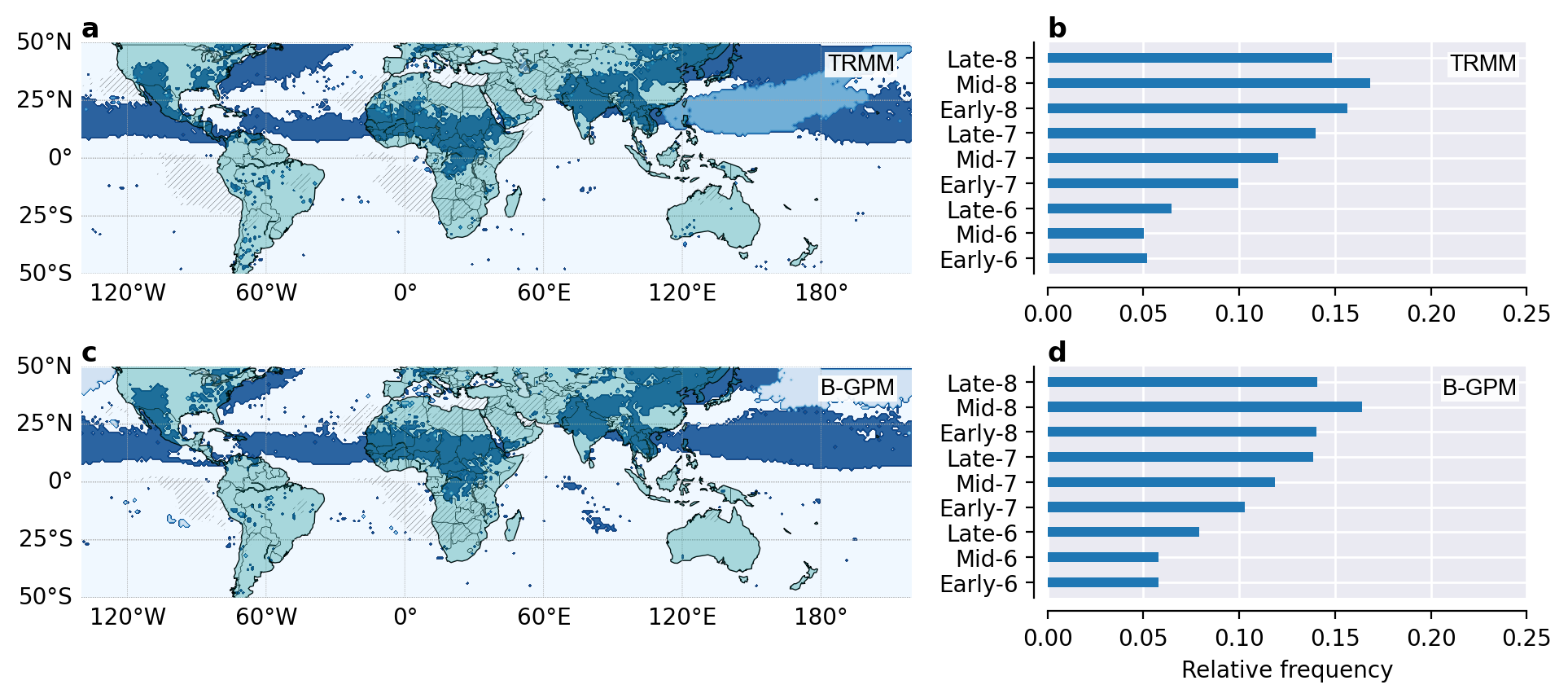}
    \caption{Same as~\hyperref[TRMM_GPMAST_CPDS_ES_2000_To_2019_JJA_900_d10_P995_1X1_T18G2]{Fig.}~\ref{TRMM_GPMAST_CPDS_ES_2000_To_2019_JJA_900_d10_P995_1X1_T18G2}, but for a different $\tau_{max}=30$ days. Communities of mutual correspondence with a similarity of $RCR($TRMM=\{2, 6\}, B-GPM=\{2, 9\}$)=0.67$ from~\hyperref[TRMM_GPMAST_LR_d30_P995]{Fig.}~\ref{TRMM_GPMAST_LR_d30_P995}.
    }
    \label{TRMM_GPMAST_CPDS_ES_2000_To_2019_JJA_900_d30_P995_1X1_T26T29}
\end{figure*}

\clearpage
\section{THE ROBUSTNESS TEST ON DIFFERENT SIGNIFICANCE LEVELS}\label[Appendix]{THE ROBUSTNESS TEST ON DIFFERENT SIGNIFICANCE LEVELS}
\begin{figure*}[!htbp]
    \centering
    \includegraphics[width=.85\textwidth]{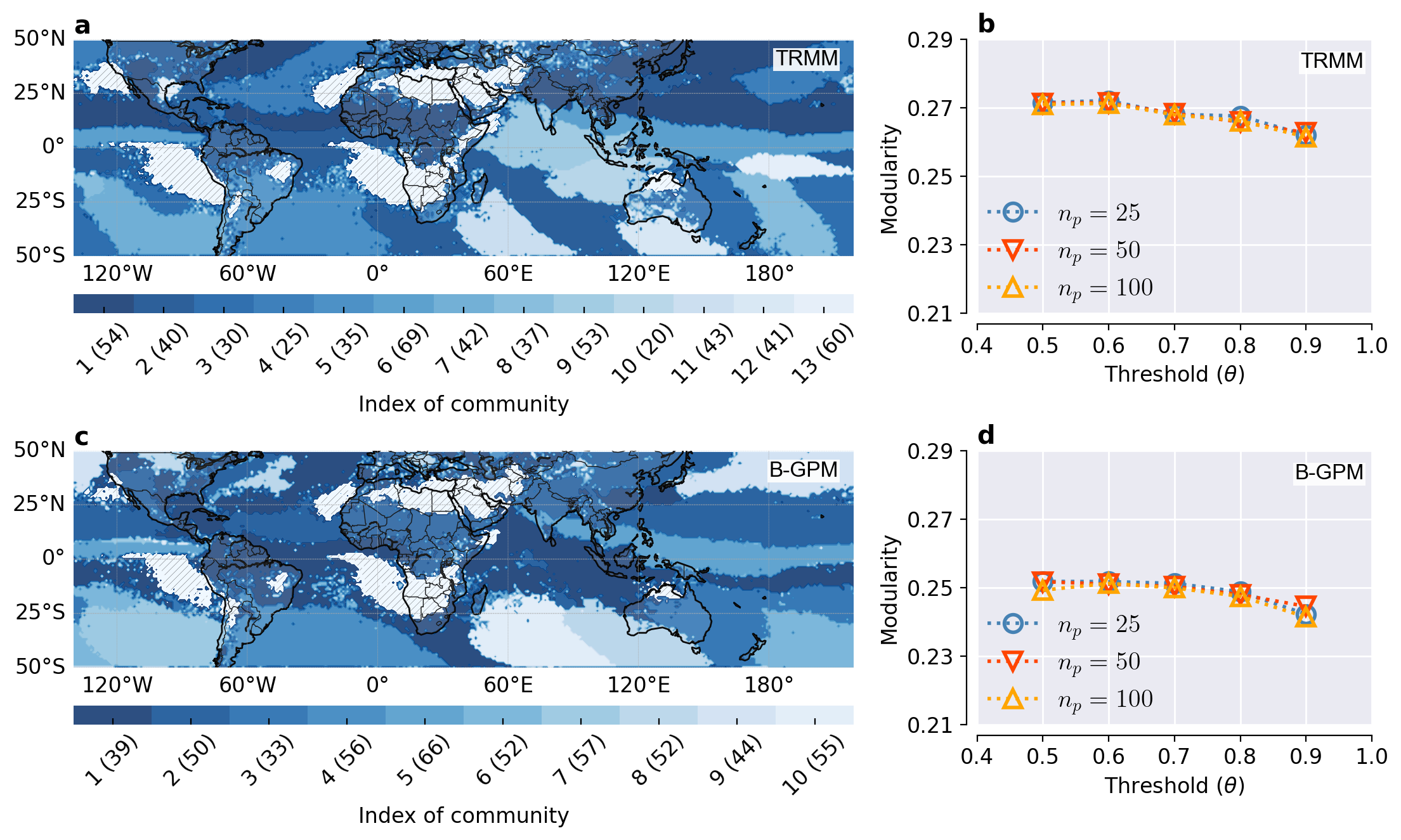}
    \caption{
        Same as~\hyperref[TRMM_GPMAST_LR_d10_P995]{Fig.}~\ref{TRMM_GPMAST_LR_d10_P995}, but for a different significance level $P < 0.006$.
        In (a) and (c), community structures for TRMM and B-GPM, are chosen according to (b) with $n_p=25$ and $\theta=0.6$ and (d) with $n_p=25$ and $\theta=0.5$, respectively, to indicate the largest modularity value.
        They share a similarity of 0.57, based on the normalized mutual information score.
    }
    \label{TRMM_GPMAST_LR_d10_P994}
\end{figure*}

\begin{figure*}[!htbp]
    \centering
    \includegraphics[width=.85\textwidth]{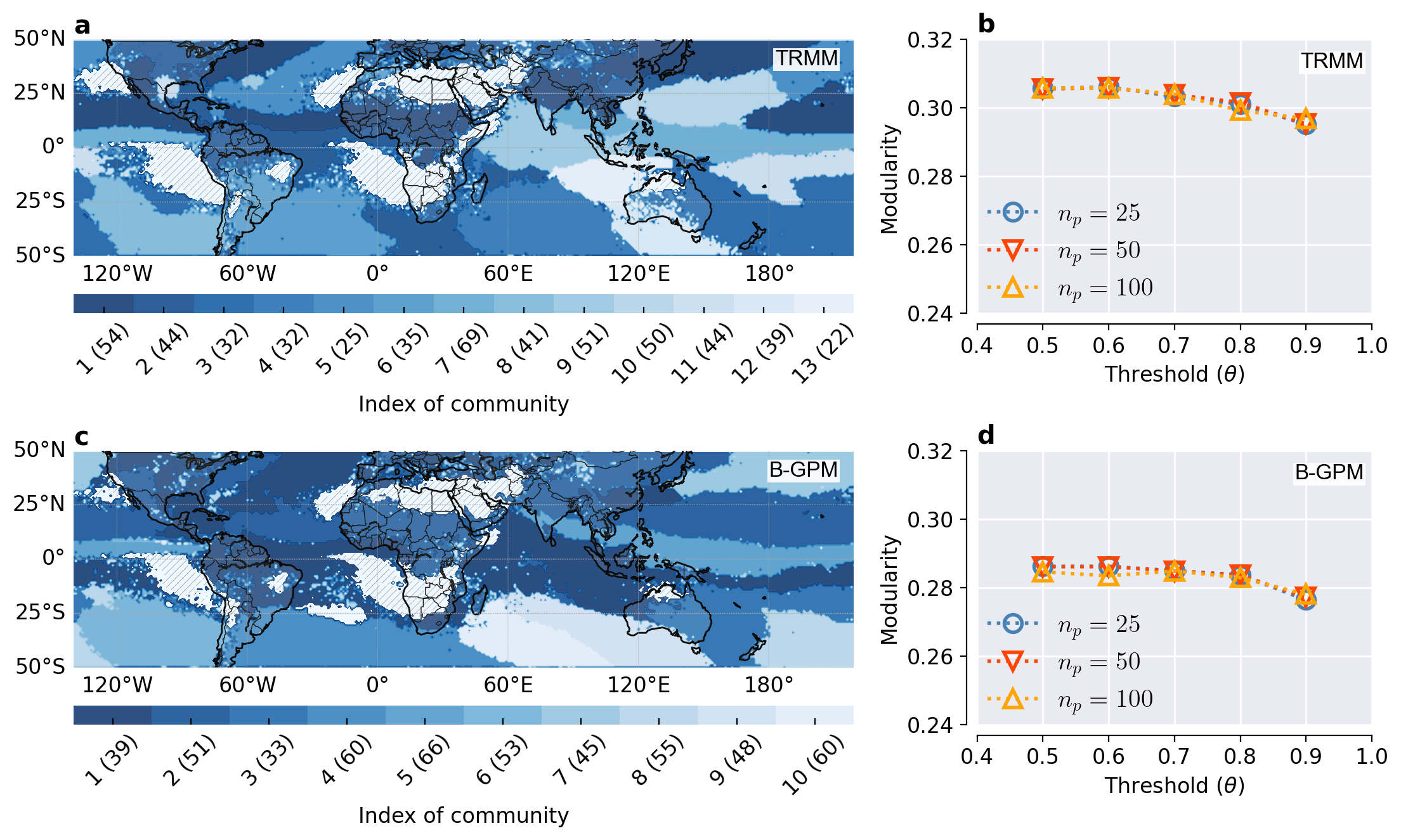}
    \caption{
        Same as~\hyperref[TRMM_GPMAST_LR_d10_P995]{Fig.}~\ref{TRMM_GPMAST_LR_d10_P995}, but for a different significance level $P<0.004$.
        Community structures for (a) TRMM and (c) B-GPM, are chosen according to (b) with $n_p=25$ and $\theta=0.6$ and (d) with $n_p=25$ and $\theta=0.6$, respectively, sharing a similarity of 0.59.
    }
    \label{TRMM_GPMAST_LR_d10_P996}
\end{figure*}

\begin{figure*}[!htbp]
    \centering
    \includegraphics[width=.95\textwidth]{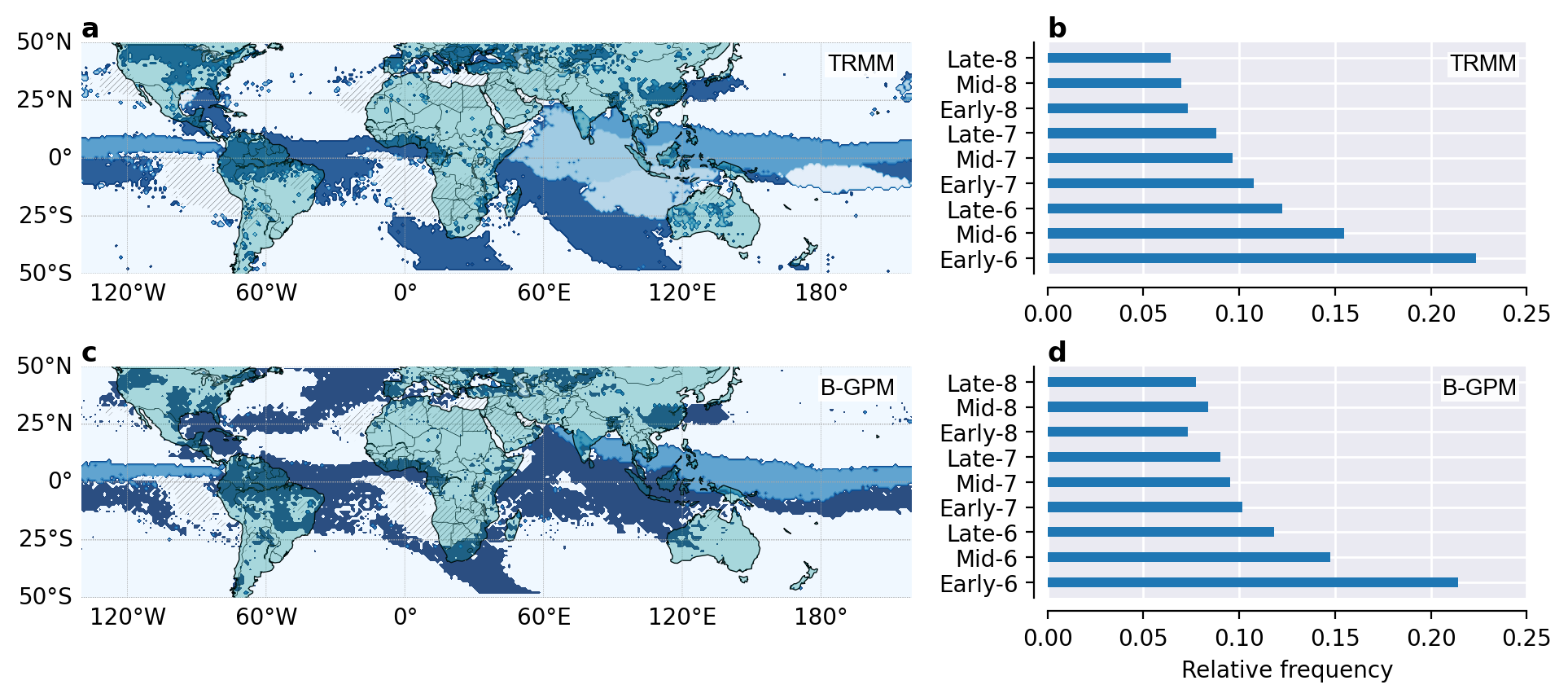}
    \caption{Same as~\hyperref[TRMM_GPMAST_CPDS_ES_2000_To_2019_JJA_900_d10_P995_1X1_T2691013G15]{Fig.}~\ref{TRMM_GPMAST_CPDS_ES_2000_To_2019_JJA_900_d10_P995_1X1_T2691013G15}, but for a different significance level $P<0.006$. Communities of mutual correspondence with a similarity of $RCR($TRMM=\{2, 6, 9, 10, 13\}, B-GPM=\{1, 5\}$)=0.53$. (a) Communities 2, 6, 9, 10 and 13 from~\hyperref[TRMM_GPMAST_LR_d10_P994]{Fig.}~\ref{TRMM_GPMAST_LR_d10_P994}\hyperref[TRMM_GPMAST_LR_d10_P994]{(a)}. (c) Communities 1 and 5 from~\hyperref[TRMM_GPMAST_LR_d10_P994]{Fig.}~\ref{TRMM_GPMAST_LR_d10_P994}\hyperref[TRMM_GPMAST_LR_d10_P994]{(b)}. (b) and (d): Frequency of synchronized days within meta-communities in (a) and (c), respectively (see Sec.~\ref{The network-based clustering workflow} `Identification of synchronized days for meta-communities'). Grid points with less than three events are excluded when extracting EREs and visualized as hatched areas here.
    }
    \label{TRMM_GPMAST_CPDS_ES_2000_To_2019_JJA_900_d10_P994_1X1_T2691013G15}
\end{figure*}

\begin{figure*}[!htbp]
    \centering
    \includegraphics[width=.95\textwidth]{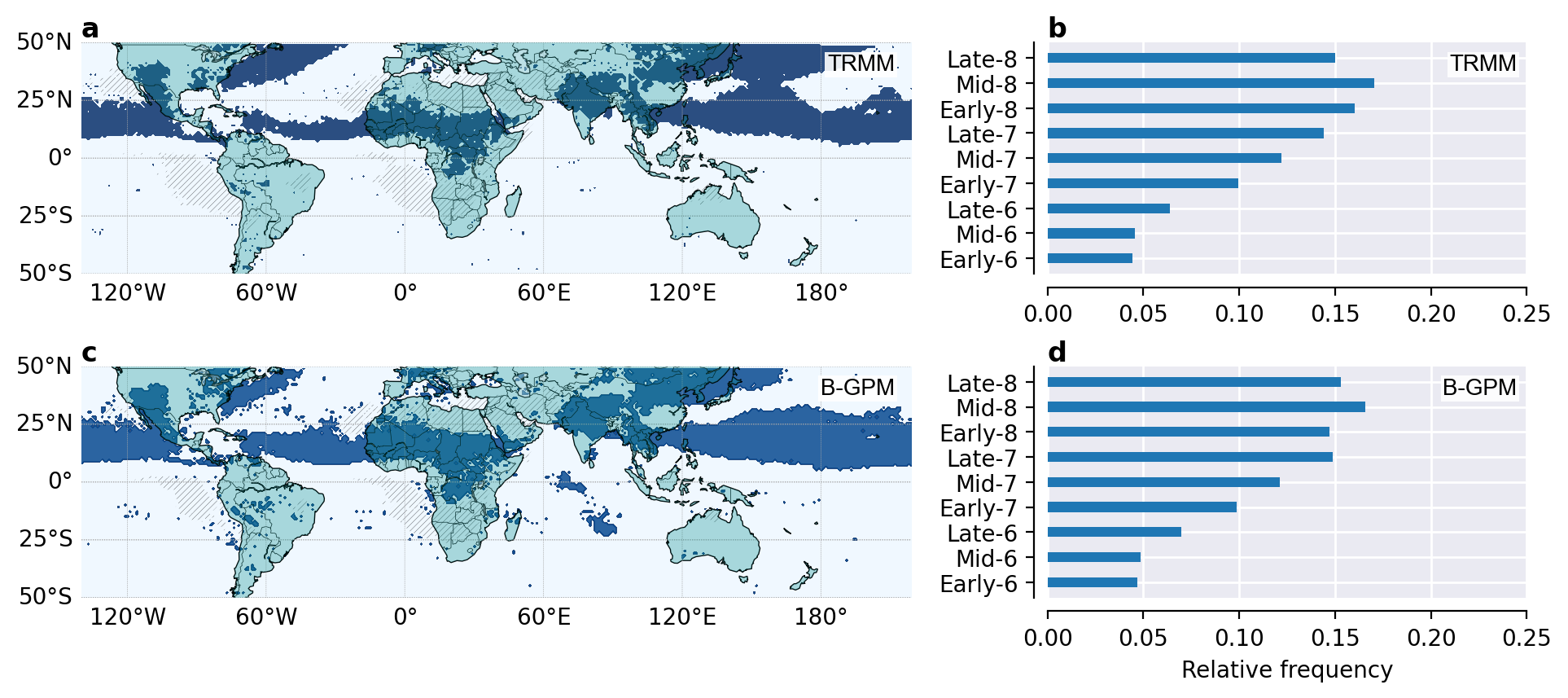}
    \caption{Same as~\hyperref[TRMM_GPMAST_CPDS_ES_2000_To_2019_JJA_900_d10_P995_1X1_T18G2]{Fig.}~\ref{TRMM_GPMAST_CPDS_ES_2000_To_2019_JJA_900_d10_P995_1X1_T18G2}, but for a different significance level $P<0.006$. Communities of mutual correspondence with a similarity of $RCR($TRMM=\{1\}, B-GPM=\{2\}$)=0.61$ from~\hyperref[TRMM_GPMAST_LR_d10_P994]{Fig.}~\ref{TRMM_GPMAST_LR_d10_P994}.
    }
    \label{TRMM_GPMAST_CPDS_ES_2000_To_2019_JJA_900_d10_P994_1X1_T1G2}
\end{figure*}

\begin{figure*}[!htbp]
    \centering
    \includegraphics[width=.95\textwidth]{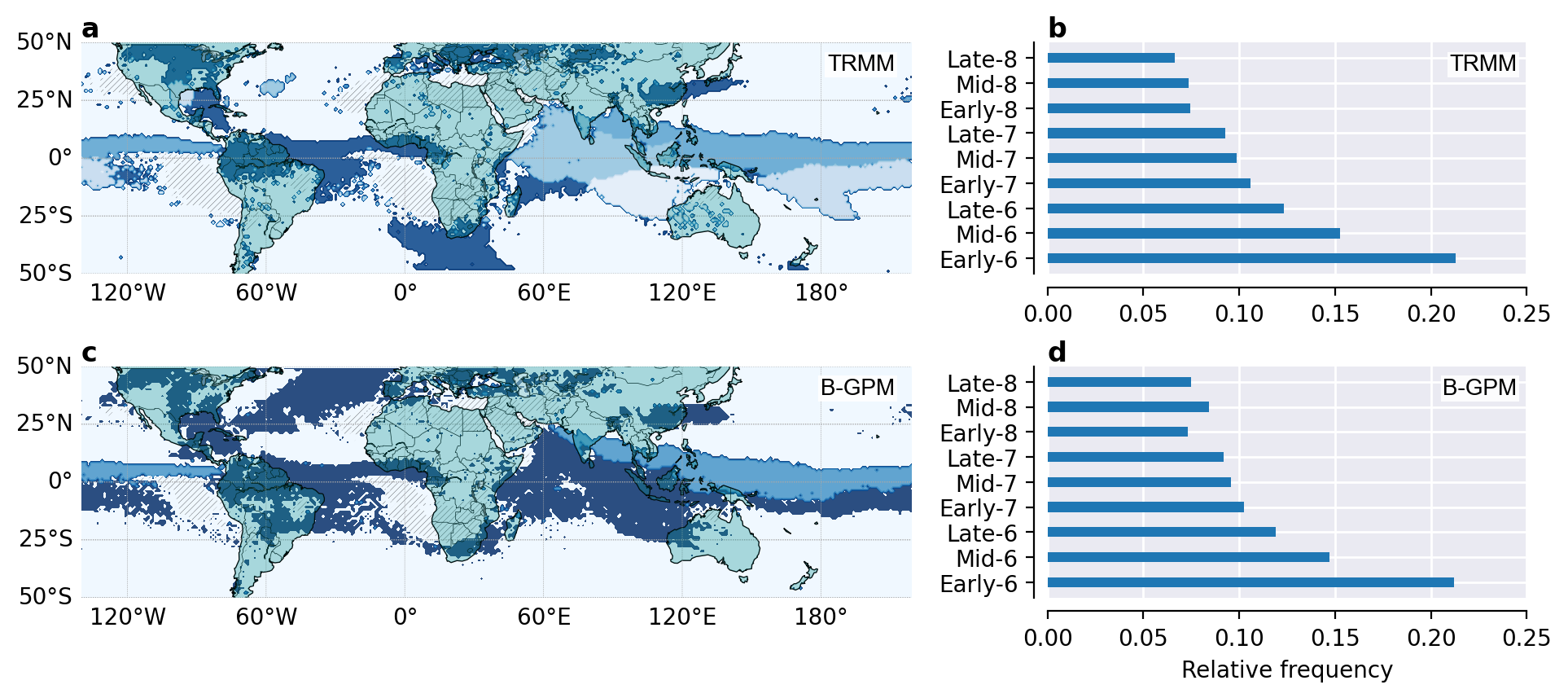}
    \caption{Same as~\hyperref[TRMM_GPMAST_CPDS_ES_2000_To_2019_JJA_900_d10_P995_1X1_T2691013G15]{Fig.}~\ref{TRMM_GPMAST_CPDS_ES_2000_To_2019_JJA_900_d10_P995_1X1_T2691013G15}, but for a different significance level $P<0.004$. Communities of mutual correspondence with a similarity of $RCR($TRMM=\{2, 7, 9, 11, 13\}, B-GPM=\{1, 5\}$)=0.57$. (a) Communities 2, 7, 9, 11 and 13 from~\hyperref[TRMM_GPMAST_LR_d10_P996]{Fig.}~\ref{TRMM_GPMAST_LR_d10_P996}\hyperref[TRMM_GPMAST_LR_d10_P996]{(a)}. (c) Communities 1 and 5 from~\hyperref[TRMM_GPMAST_LR_d10_P996]{Fig.}~\ref{TRMM_GPMAST_LR_d10_P996}\hyperref[TRMM_GPMAST_LR_d10_P996]{(b)}. (b) and (d): Frequency of synchronized days within meta-communities in (a) and (c), respectively (see Sec.~\ref{The network-based clustering workflow} `Identification of synchronized days for meta-communities'). Grid points with less than three events are excluded when extracting EREs and visualized as hatched areas here.
    }
    \label{TRMM_GPMAST_CPDS_ES_2000_To_2019_JJA_900_d10_P996_1X1_T2791113G15}
\end{figure*}

\begin{figure*}[!htbp]
    \centering
    \includegraphics[width=.95\textwidth]{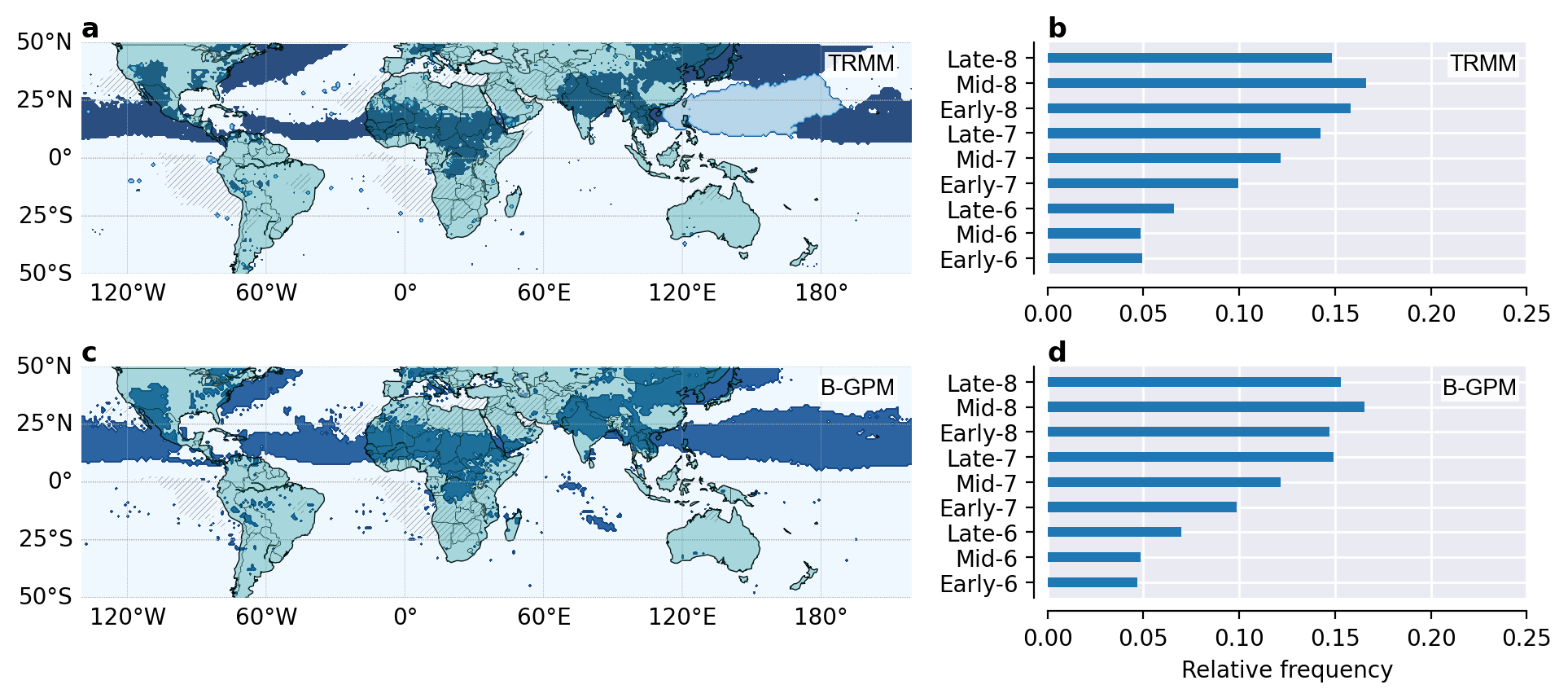}
    \caption{Same as~\hyperref[TRMM_GPMAST_CPDS_ES_2000_To_2019_JJA_900_d10_P995_1X1_T18G2]{Fig.}~\ref{TRMM_GPMAST_CPDS_ES_2000_To_2019_JJA_900_d10_P995_1X1_T18G2}, but for a different significance level $P<0.004$. Communities of mutual correspondence with a similarity of $RCR($TRMM=\{1, 10\}, B-GPM=\{2\}$)=0.64$ from~\hyperref[TRMM_GPMAST_LR_d10_P996]{Fig.}~\ref{TRMM_GPMAST_LR_d10_P996}.
    }
    \label{TRMM_GPMAST_CPDS_ES_2000_To_2019_JJA_900_d10_P996_1X1_T110G2}
\end{figure*}


\clearpage
\twocolumngrid
\section*{References}
\bibliography{Clustering_Chaos}
\end{document}